\documentclass[11pt]{article}

\usepackage[a4paper,left=16mm,right=16mm,top=22mm,bottom=22mm]{geometry}
\usepackage{amsmath,amssymb,bm}
\usepackage{graphicx}
\usepackage{booktabs}
\usepackage{microtype}
\usepackage{placeins}
\usepackage[numbers,sort&compress]{natbib}
\usepackage[font=small,labelfont=bf]{caption}
\usepackage[hidelinks]{hyperref}

\renewcommand{\topfraction}{0.9}
\renewcommand{\textfraction}{0.08}
\renewcommand{\floatpagefraction}{0.8}

\graphicspath{{figures/}}

\title{Reservoir-conditioned virtual returns generate random Liouvillian skin localization in a reciprocal Mott insulator}

\author{\parbox{0.94\textwidth}{\centering
Y. T. Wang$^{1}$, X. Z. Zhang$^{1,2,*}$, and W. M. Liu$^{3}$\\[5pt]
\small $^{1}$College of Physics and Materials Science, Tianjin Normal University, Tianjin 300387, China\\
\small $^{2}$Interdisciplinary Center, Tianjin Normal University, Tianjin 300387, China\\
\small $^{3}$Beijing National Laboratory for Condensed Matter Physics,\\
\small Institute of Physics, Chinese Academy of Sciences, Beijing 100190, China\\[4pt]
\small $^{*}$email: zhangxz@tjnu.edu.cn}}
\date{}

\begin{document}
\maketitle

\begin{abstract}
Directional dissipation can concentrate relaxation modes in space, but connecting this accumulation to controlled microscopic processes in correlated matter requires separating virtual charge motion from spin coherence.
We study this connection in a half-filled Hubbard chain with reciprocal hopping and number-conserving, direction-selective returns of virtual doublon--hole defects.
Eliminating charge defects and then Mott coherences yields asymmetric spin-exchange rates under an explicit separation of timescales, with full short-chain dynamics supporting the reduction on finite exchange times.
Balanced random returns concentrate the stationary population and several low-lying right modes at sample-selected positions rather than a predetermined edge, producing random Liouvillian skin localization despite zero end-to-end logarithmic rate bias.
The accumulated rate imbalance determines an exact finite-density hard-core stationary state and a one-particle activated relaxation scale governed by random barriers.
This connection between locally calibrated return rates, many-body stationary weights and slow relaxation offers a means of controlling transport in constrained open quantum matter without changing the reciprocal Hamiltonian.
\end{abstract}

\section*{Introduction}

\begin{figure}[t]
\centering
\includegraphics[width=17.8cm]{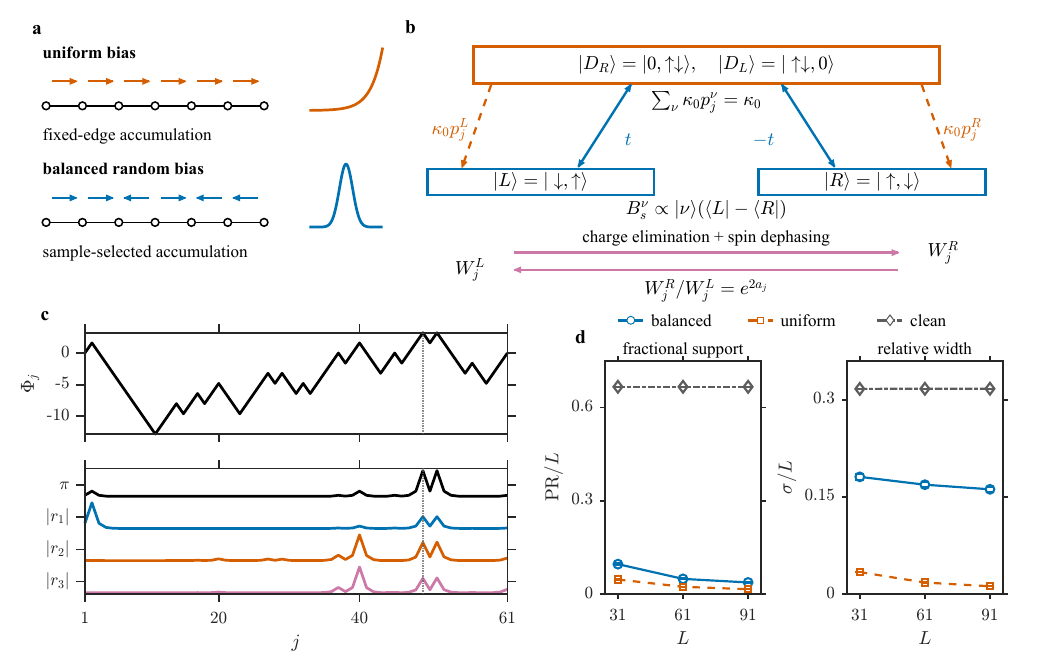}
\caption{\textbf{Virtual-state returns generate directional spin exchange and internal low-mode localization.}
\textbf{a}, Uniform and balanced random bond preferences illustrate fixed-edge and sample-selected accumulation. Arrows indicate the locally preferred exchange direction; the profile sketches are schematic.
\textbf{b}, The grouped charge box contains two oriented source defects whose reservoir returns are environmentally distinguishable. Each defect is coupled coherently to both Mott states with amplitudes \(t,-t\), and returns to either state with partial rate \(\kappa_0p_j^{L,R}\). The effective jumps retain both amplitudes. Only after Mott coherence is suppressed do they yield the directional population rates shown below.
\textbf{c}, Balanced sample with \(L=61\), \(a=0.8\), selected by proximity to the ensemble-median participation and width, not by peak alignment (Methods). The stationary vector \(\pi\) and first three nonstationary physical right-mode magnitudes are individually maximum-normalized and vertically offset. The dashed line marks the first global maximum of \(\Phi_j\); different modes need not peak there.
\textbf{d}, Fractional support \(\mathrm{PR}/L\) and relative width \(\sigma/L\), averaged over the first four nonstationary modes within each sample and then over 80 independent balanced samples at each \(L=31,61,91\). Bars are standard errors across samples. Uniform bias \(a_j=0.8\) and clean rates are deterministic controls. All numerical panels use reflecting boundaries; Supplementary Fig.~4 gives complementary peak diagnostics.}
\label{fig:overview}
\end{figure}

Nonreciprocal hopping changes localization and boundary sensitivity~\cite{HatanoNelson1996,HatanoNelson1997}; in lattice systems it can concentrate right eigenmodes at an open edge, producing the non-Hermitian skin effect~\cite{YaoWang2018,Kunst2018,Okuma2020}. This connects spatial accumulation to the broader study of non-Hermitian spectra and topology~\cite{Ashida2020,Bergholtz2021,Zhang2022Review}. Directional dissipation can likewise reshape decay modes and relaxation~\cite{Song2019Chiral,Longhi2020,Haga2021,Huang2024NonHermitian}. A uniform bias fixes the preferred edge. Balanced random return preferences remove this systematic endpoint preference while retaining local directional fluctuations (Fig.~\ref{fig:overview}a). Erratic Hamiltonian and Liouvillian models already exhibit sample-dependent internal localization~\cite{Longhi2025ENHSL,Longhi2026QST}; the latter also connects prescribed random rates to Sinai-type transport~\cite{Sinai1982,BouchaudGeorges1990}. For correlated matter, the question is how to derive these rates from controlled virtual processes and when they remain predictive for many constrained excitations.

For single-particle bands, anomalous edge states and the failure of conventional bulk--boundary correspondence motivated non-Bloch descriptions~\cite{Lee2016,Xiong2018,Yokomizo2019}, alongside symmetry-based classifications~\cite{Gong2018,Kawabata2019}. Periodic driving~\cite{Zhang2020NonHermitian}, geometric dualities~\cite{Zhou2022Curving} and higher-dimensional boundary spectra~\cite{WVincent2022,ZhaoErhai2023} further show that spatial response depends on more than a preferred hopping direction. Electrical-circuit and photonic experiments make these boundary effects accessible~\cite{Helbig2020,Weidemann2020,Xiao2020}, while dynamical skin effects emphasize their time-dependent signatures~\cite{Li2022Dynamic}. In an interacting open system, however, the motion of the retained excitations must be derived together with the dissipative processes that create and destroy coherence.

Interactions already modify skin localization through the spectral structure of the interacting Hatano--Nelson model~\cite{Zhang2022Symmetry}, nonreciprocal motion of repulsively bound pairs~\cite{Brighi2024Nonreciprocal}, and virtual bound-particle motion in reciprocal dissipative lattices~\cite{Li2026Dissipation}. Mott and Fock-space constraints produce further forms of accumulation~\cite{Yoshida2024Mott,Shimomura2024Fock}, including a connection between Fock-space concentration and quantum scars~\cite{Shen2024}. Two-body loss can generate many-body Liouvillian skin dynamics~\cite{Hamanaka2023}, and exact steady states are known in dissipative skin models~\cite{Yang2022,Hu2025Exact}. The question here is therefore not whether interactions or reservoirs can produce nonreciprocity. We ask how a number-conserving return process transfers directional information from virtual charge defects to Mott spins, and how the resulting rate ratios determine their finite-density stationary distribution when the end-to-end bias vanishes.

At half filling and large repulsion, reciprocal Hubbard hopping connects singly occupied spin states through virtual doublon--hole configurations~\cite{Hubbard1963}. In a closed chain these paths recombine into reciprocal Hermitian superexchange~\cite{Anderson1950}. Optical-lattice Hubbard systems provide a setting in which the charge and spin scales can be controlled separately~\cite{JakschZoller2005,Bloch2008}. Direction-selective decay alone is insufficient to obtain classical exchange rates: elastic return and spin exchange remain coherent within each source--final-state channel, and changing a return rate can also change the defect lifetime. We use fixed total linewidths to vary the final-state probabilities independently, then suppress the remaining spin coherence on an intermediate timescale. Both eliminations are controlled by an explicit hierarchy~\cite{Reiter2012,Kessler2012}. Unequal source lifetimes preserve the directional ratio only if all sources share the same final-state probabilities and both Mott states couple with equal magnitude to each source.

We derive the Mott population generator and compare both reductions with full number-conserving Hubbard--Lindblad dynamics on finite exchange times. Balanced random returns concentrate the stationary population and the first few physical right modes at sample-selected positions along the chain. Their rate ratios fix the stationary measure in every magnetization sector, while the common bond activities also enter the relaxation scale. For one excitation, a path bound relates the logarithmic population gap to a Brownian-bridge barrier up to logarithmic corrections; few-particle spectra support the corresponding dilute trend. A proposed cold-atom protocol reconstructs these rates from local measurements before comparing them with many-body profiles. The central link is thus between a measurable virtual-state return process and both the spatial organization and slow relaxation of the retained spins.

\section*{Results}

\subsection*{Reservoir-conditioned virtual returns generate asymmetric Mott exchange}

The half-filled repulsive Hubbard chain separates low-energy spin motion from virtual charge excitation. Each site is singly occupied in the Mott manifold, and hopping creates a doublon--hole pair at an energy cost set by the repulsion. Because the closed chain produces reciprocal superexchange, leaving its hopping unchanged identifies the reservoir return as the source of directional spin motion (Fig.~\ref{fig:overview}b). This differs from Hubbard models whose non-Hermitian Hamiltonians support pairing ground states or modified superfluid phases~\cite{ZhangXz2021,Takemori2024}. We use two-component fermions on an open chain at half filling, with \(\hbar=1\),
\begin{equation}
H_{\mathrm{Hub}}=-t\sum_{j=1}^{L-1}\sum_{\sigma=\uparrow,\downarrow}
\left(c_{j+1,\sigma}^{\dagger}c_{j,\sigma}+\mathrm{H.c.}\right)
+U\sum_{j=1}^{L}n_{j,\uparrow}n_{j,\downarrow}.
\label{eq:Hubbard}
\end{equation}
Here \(c_{j,\sigma}\) annihilates a fermion, \(n_{j,\sigma}=c_{j,\sigma}^{\dagger}c_{j,\sigma}\), \(t>0\) is reciprocal hopping and \(U>0\) is the repulsion. We denote the hopping term by \(H_t\) and work at \(U/t\gg1\). The low-energy projector \(P\) retains one atom per site; \(Q=1-P\) contains charge defects. We count down spins relative to the fully up-polarized reference, so that \(\eta_j=1/2-S_j^z\) and \(M=\sum_j\eta_j\). On bond \(j\), \(|L_j\rangle=|\downarrow,\uparrow\rangle\) and \(|R_j\rangle=|\uparrow,\downarrow\rangle\); the move \(L_j\to R_j\) is a rightward down-spin exchange. Both states couple to \(|D_{j,R}\rangle=|0,\uparrow\downarrow\rangle\) and \(|D_{j,L}\rangle=|\uparrow\downarrow,0\rangle\). Their hopping amplitudes have equal magnitudes and opposite signs in the Fock convention specified in Methods.

For each source defect \(s=R,L\), two environmentally distinguishable reservoir channels return to final state \(\nu=R,L\):
\begin{equation}
R_{j,s}^{\nu}=\sqrt{\kappa_0p_j^{\nu}}\,|\nu_j\rangle\langle D_{j,s}|,
\qquad p_j^{R,L}=\frac{e^{\pm a_j}}{2\cosh a_j}.
\label{eq:reset}
\end{equation}
Spectator sites are unchanged. Each reset conserves atom number and total magnetization. Since \(p_j^R+p_j^L=1\), every source defect has total decay rate \(\kappa_0\). The final-state preference \(a_j\) enters the dissipator, not Eq.~\eqref{eq:Hubbard}. Figure~\ref{fig:overview}b shows this return network; the complete master equation, including spin monitoring, is Eq.~\eqref{eq:master}. Thus the Hamiltonian is reciprocal, whereas the full open-system dynamics is direction-selective by construction.

Charge elimination alone retains the interference between the two initial Mott states. On an isolated bond its effective jumps and Hamiltonian are
\begin{equation}
B_{j,s}^{\nu}=-\frac{t\sqrt{\kappa_0p_j^{\nu}}}{U-i\kappa_0/2}
|\nu_j\rangle(\langle L_j|-\langle R_j|),\qquad
H_{\mathrm{ex},j}=-\frac{2Ut^2}{D_0}(|L_j\rangle-|R_j\rangle)(\langle L_j|-\langle R_j|),
\label{eq:coherent_effective}
\end{equation}
where \(D_0=U^2+(\kappa_0/2)^2\). Overall phases of individual jumps are immaterial. The two sources are separate dissipators, but the two amplitudes within each \(B_{j,s}^{\nu}\) cannot be separated without removing coherence. In particular, the triplet \((|L_j\rangle+|R_j\rangle)/\sqrt2\) is dark without monitoring.

We resolve Mott configurations with dephasing rate \(\gamma_{\mathrm m}\). For configuration-projector monitoring, Mott--Mott coherences decay at \(\gamma_{\mathrm m}\), and Mott--charge coherences acquire an additional width \(\gamma_{\mathrm m}/2\). A sufficient local hierarchy for the two-stage reduction is
\begin{equation}
\frac{t}{\sqrt{D_0}}\ll1,\quad
\frac{\gamma_{\mathrm m}}{\kappa_0}\ll1,\quad
\frac{W_{\max}}{\gamma_{\mathrm m}}\ll1,\quad
\frac{(J_{\mathrm{rec}}+W_{\max})^2}{\gamma_{\mathrm m}W_{\min}}\ll1,
\label{eq:hierarchy}
\end{equation}
where \(J_{\mathrm{rec}}=2Ut^2/D_0\) is the off-diagonal pair-exchange matrix element; the conventional Heisenberg coupling is \(2J_{\mathrm{rec}}\). Here \(W_{\min},W_{\max}\) are the smallest and largest target bond rates. These conditions are compatible at fixed \(U,\kappa_0\) and bounded \(a_j\): taking \(t\to0\) with \(\gamma_{\mathrm m}=t\) makes each ratio vanish. The monitoring scale is therefore intermediate: it is slow compared with the charge decay but fast compared with the effective Mott dynamics. After this second elimination, the population rates are
\begin{equation}
W_j^{R,L}=\Gamma_j e^{\pm a_j},\qquad
\Gamma_j=\frac{\kappa_0t^2}{\cosh a_j\,D_0}.
\label{eq:rates}
\end{equation}
For binary \(a_j=\pm a\), \(\Gamma_j=\Gamma\). The common linewidth makes the exchange scale explicit. More generally, unequal source lifetimes change the common bond activity without changing \(W_j^R/W_j^L\) when all sources share the same final-state probabilities and the two Mott states couple with equal magnitude to each source (Supplementary Note~1).

\subsection*{Full Hubbard--Lindblad dynamics benchmarks the controlled population reduction}

To test the reduction on the exchange timescale, we retain all charge configurations in the full number-conserving Hubbard--Lindblad equation. The fixed-\((N,M)\) Hilbert spaces have dimensions 4 for \(L=2,M=1\) and 9 for \(L=3,M=1\). We compare their evolution with the charge-eliminated model in Eq.~\eqref{eq:coherent_effective}, including its elastic--exchange cross terms, and with \(G^{(1)}\). The initial down spin occupies the leftmost site. The rates in Eq.~\eqref{eq:rates} fix the time axis in all three calculations; no rate rescaling is fitted.

\begin{figure}[!htbp]
\centering
\includegraphics[width=\textwidth]{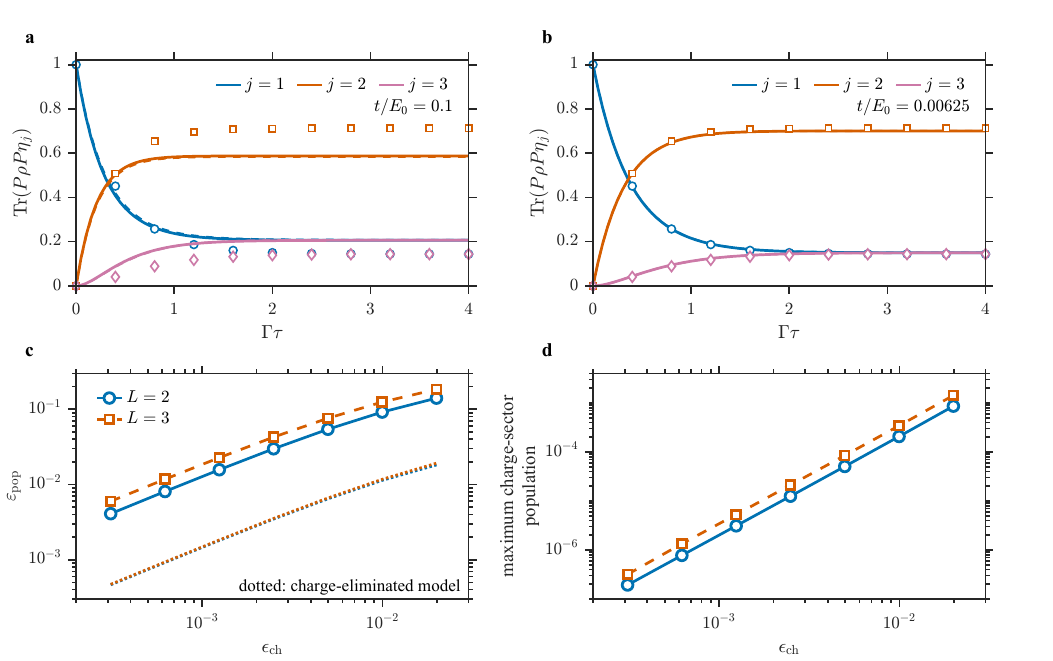}
\caption{\textbf{Full short-chain dynamics benchmarks the two controlled reductions on exchange times.}
\textbf{a,b}, Full \(L=3,M=1\) dynamics (solid), charge-eliminated density-matrix model (dashed), and population generator (markers), with \(U/E_0=10\), \(\kappa_0/E_0=2\), \(\gamma_{\mathrm m}=t\) and \((a_1,a_2)=(0.8,-0.8)\), at the indicated hopping. Initially the down spin occupies site 1 and all sites are singly occupied. Colors label sites, and \(\tau\) denotes time.
\textbf{c}, Maximum absolute configuration-population difference \(\varepsilon_{\mathrm{pop}}\) from the full model on 81 uniformly spaced times \(0\le\Gamma\tau\le4\), plotted against \(\epsilon_{\mathrm{ch}}=t/\sqrt{D_0}\). Solid/dashed marker curves compare \(G\) for \(L=2/3\); dotted curves compare the coherent effective model. The two-site bond has \(a_1=0.8\) and is not balanced.
\textbf{d}, Maximum charge-sector population on that grid. These are deterministic small-system comparisons, without disorder error bars or fitted rate rescaling; projected Mott populations are unnormalized. The comparison is restricted to the displayed short chains and finite exchange-time interval.}
\label{fig:projection}
\end{figure}

Figure~\ref{fig:projection}a,b compares trajectories at \(U=10E_0\), \(\kappa_0=2E_0\) and \(\gamma_{\mathrm m}=t\), where \(E_0\) is a reference energy. Lowering \(t/E_0\) reduces monitoring-induced charge broadening and feedback from Mott coherence, as measured by the population differences in Fig.~\ref{fig:projection}c, and reduces charge admixture (Fig.~\ref{fig:projection}d). The comparison uses the unnormalized block \(P\rho P\); normalization cannot hide leakage. Both reduced models approach the microscopic trajectories over the displayed exchange-time interval, with the coherent effective model closer before the population limit is reached. Separated defects in the three-site model are included. These calculations benchmark the two local reductions over the displayed finite exchange-time interval for short chains; they do not establish a length-uniform reduction, the complete long-chain Liouvillian spectrum or convergence on exponentially long activated times.

\subsection*{Balanced return ratios produce sample-dependent internal localization}

The retained configuration \(\boldsymbol\eta=(\eta_1,\ldots,\eta_L)\) has \(\eta_j\in\{0,1\}\) and \(\sum_j\eta_j=M\). A move \(10\to01\) occurs at \(W_j^R\), and its reverse at \(W_j^L\). We use a column-conserving population generator \(\dot{\bm p}=G^{(M)}\bm p\). In the transition-resolved quantum embedding, this population generator is an invariant diagonal Liouvillian block, so its right eigenvectors are genuine relaxation modes of that resolved open-system dynamics. Its one-dimensional rate structure~\cite{Derrida1983} and exclusion constraint~\cite{Liggett1999,KipnisLandim1999} are familiar; the construction above supplies its microscopic rates and the conditions for obtaining a closed population dynamics.

For an odd chain length, balanced binary disorder obeys
\begin{equation}
\begin{aligned}
a_j&\in\{+a,-a\},\quad \sum_{j=1}^{L-1}a_j=0,\quad
\Phi_1=0,\quad \Phi_{j+1}-\Phi_j=\log\frac{W_j^R}{W_j^L}=2a_j,\\
\mathcal B_{\mathrm{end}}&\equiv\sum_{j=1}^{L-1}\log\frac{W_j^R}{W_j^L}
=\Phi_L-\Phi_1=0.
\end{aligned}
\label{eq:phi}
\end{equation}
The accumulated bias is therefore a random-walk bridge. We refer to \(\mathcal B_{\mathrm{end}}=0\) as zero end-to-end logarithmic rate bias. This is not a thermodynamic cycle affinity because the open chain contains no nontrivial spatial cycle. The balanced condition removes systematic endpoint preference while retaining the internal random landscape, in contrast to the uniform-bias case in Fig.~\ref{fig:overview}a. It does not imply that the instantaneous local drift vanishes; any finite reflecting birth--death chain has zero stationary current.

For one excitation, detailed balance gives \(\pi_j\propto e^{\Phi_j}\). Following the random-rate picture~\cite{Longhi2026QST}, we use random Liouvillian skin localization for sample-dependent concentration of physical population modes despite zero end-to-end logarithmic rate bias. Our numerical evidence concerns the stationary population and first four nonstationary modes. Finite samples may have endpoint maxima or ties. The sample in Fig.~\ref{fig:overview}c is selected by its participation and width, without using peak alignment with the landscape maximum. The ensemble data in Fig.~\ref{fig:overview}d distinguish these modes from clean extended modes and uniformly biased boundary accumulation. Over \(L=31,61,91\), they occupy a decreasing fraction of the chain, while separated internal peaks can retain a finite relative width. This does not establish single-center exponential localization. Supplementary Fig.~4a,b records peak locations and endpoint exceptions.

For one excitation, let \(D=\operatorname{diag}(\pi_1,\ldots,\pi_L)\). Detailed balance makes \(A=-D^{-1/2}GD^{1/2}\) symmetric. If \(Au_q=\lambda_q u_q\), then the physical right mode \(r_q=D^{1/2}u_q\) satisfies \(Gr_q=-\lambda_q r_q\). Reversibility therefore coexists with spatial reweighting in the physical basis. Mode accumulation and point-gap topology require separate analysis, as anomalous skin models also illustrate~\cite{Guo2023Anomalous}. Here the similarity transformation explains the reweighting; our low-mode data imply neither a point-gap invariant nor localization of an extensive spectral fraction.

\subsection*{Hard-core exclusion gives an exact finite-density stationary state}

The same ratio determines the unique stationary state for every fixed \(M\):
\begin{equation}
P_{\mathrm{ss}}^{(M)}(\boldsymbol\eta)=\frac{1}{Z_M}
\exp\!\left(\sum_j\eta_j\Phi_j\right)
\delta_{\sum_j\eta_j,M}.
\label{eq:pss}
\end{equation}
An allowed exchange changes this weight by \(e^{\Phi_{j+1}-\Phi_j}=W_j^R/W_j^L\), proving pairwise detailed balance. Positive rates make the fixed-sector graph irreducible. This is an exact canonical hard-core measure for \(G^{(M)}\), including spatially varying positive \(\Gamma_j\). It describes exclusion in the rate-generated potential, not occupation of independent one-particle eigenmodes.

\begin{figure}[!htbp]
\centering
\includegraphics[width=\textwidth]{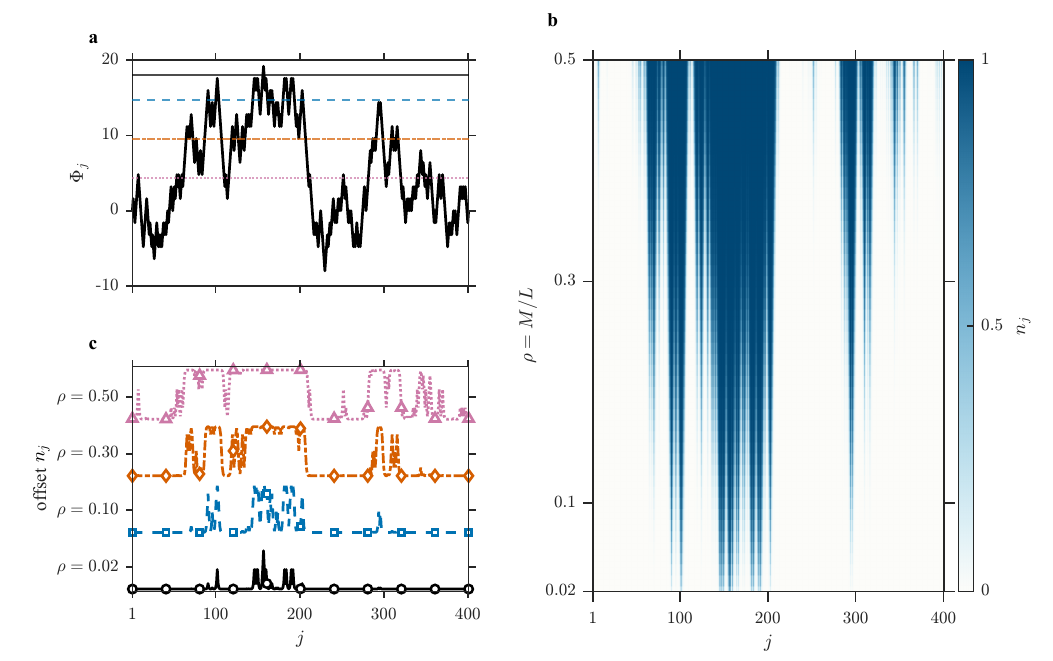}
\caption{\textbf{Hard-core exclusion converts the rate-generated landscape into an exact finite-density filling problem.}
\textbf{a}, Representative balanced profile with \(L=401\), \(a=0.8\). Lines mark the grand-canonical waterlines at \(M=8,40,120,201\); density labels are rounded.
\textbf{b}, Exact canonical density for 49 fillings between approximately 0.02 and 0.50 in the same sample.
\textbf{c}, The four corresponding exact canonical density profiles, vertically offset for legibility. All calculations use reflecting boundaries. The displayed landscape is a representative fixed realization, not a disorder average; its complete bond sequence is supplied in Source Data.}
\label{fig:filling}
\end{figure}

With weights \(z_j=e^{\Phi_j}\), \(Z_M\) is their order-\(M\) elementary symmetric polynomial. Prefix and suffix recursions give the exact site densities in \(O(LM)\) operations (Methods). Occupying a favorable site removes it from the remaining configuration space, so additional spin excitations compete for secondary extrema and their surrounding regions. The grand-canonical approximation is
\begin{equation}
n_j^{\mathrm{gc}}=\frac{1}{1+e^{-(\Phi_j+\mu)}},\qquad
\sum_jn_j^{\mathrm{gc}}=M.
\label{eq:waterline}
\end{equation}
The waterline \(-\mu\) decreases as more down spins are added (Fig.~\ref{fig:filling}a). The exact density map and line cuts show saturation of the highest-weight sites and progressive occupation of secondary regions (Fig.~\ref{fig:filling}b,c). These spin-density puddles are irregular, possibly disconnected occupation profiles imposed by exclusion, not attraction-bound droplets or a new phase. Supplementary Fig.~1a compares canonical and grand-canonical densities, and Supplementary Fig.~1b quantifies the filling geometry. Changes of positive bond activities at fixed directional ratios leave Eq.~\eqref{eq:pss} unchanged exactly, whereas accumulated ratio errors can alter the profile. Methods and Supplementary Note~6 give a finite-size bound on this sensitivity.

\subsection*{Random-walk barriers set the activated dilute relaxation scale}

We denote the smallest nonzero decay rate of the population generator by \(\Delta_G^{(M)}\). It need not equal the gap of an unrestricted density-matrix Liouvillian. For a clean reflecting chain,
\begin{equation}
\Delta_G^{(1)}=2\Gamma[1-\cos(\pi/L)]\simeq\pi^2\Gamma/L^2.
\label{eq:clean_gap}
\end{equation}
For random rates, let \(m_j=\sum_{\ell\le j}\pi_\ell\) and \(c_j=\pi_jW_j^R=\pi_{j+1}W_j^L\). The minimum spatial-cut conductance is
\begin{equation}
\phi=\min_{1\le j<L}\frac{c_j}{\min(m_j,1-m_j)}.
\label{eq:oneconductance}
\end{equation}
Conductance bounds relate relaxation to the probability flow across a bottleneck~\cite{Cheeger1970,JerrumSinclair1989,LevinPeresWilmer2009}. On a path, the spatial-cut minimum equals the minimum over all subsets. An indicator test function gives the upper bound below; summing the Dirichlet form along paths gives the lower bound (Supplementary Note~4):
\begin{equation}
\frac{\phi}{L-1}\le\Delta_G^{(1)}\le2\phi.
\label{eq:pathbound}
\end{equation}
Thus the logarithmic gap and conductance differ by at most \(O(\log L)\), sample by sample, without equating fitted finite-size slopes.

For uniform \(\Gamma\), define the dimensionless two-sided barrier
\begin{equation}
\mathcal B_L=\max_{j<L}\left[
\min\!\left(\max_{\ell\le j}\Phi_\ell,\max_{\ell>j}\Phi_\ell\right)-\Phi_j-a_j\right].
\label{eq:barrier}
\end{equation}
Replacing each partial partition sum in Eq.~\eqref{eq:oneconductance} by its largest term yields
\begin{equation}
\mathcal B_L-\log2\le-\log(\Delta_G^{(1)}/\Gamma)
\le\mathcal B_L+\log L+\log(L-1).
\label{eq:barrierbound}
\end{equation}
Balanced random-walk bridges have excursions on the scale \(a\sqrt L\); their two-sided barrier is a fluctuating functional on the same scale. Equation~\eqref{eq:barrierbound} therefore gives the one-particle Sinai-type activated scale~\cite{Sinai1982,BouchaudGeorges1990}. Before disorder averaging, the rescaled logarithmic gap remains sample dependent and converges in distribution to the Brownian-bridge barrier functional. Disorder-averaged quantities may possess a deterministic mean coefficient, but individual samples do not share a universal prefactor.

\begin{figure}[!htbp]
\centering
\includegraphics[width=\textwidth]{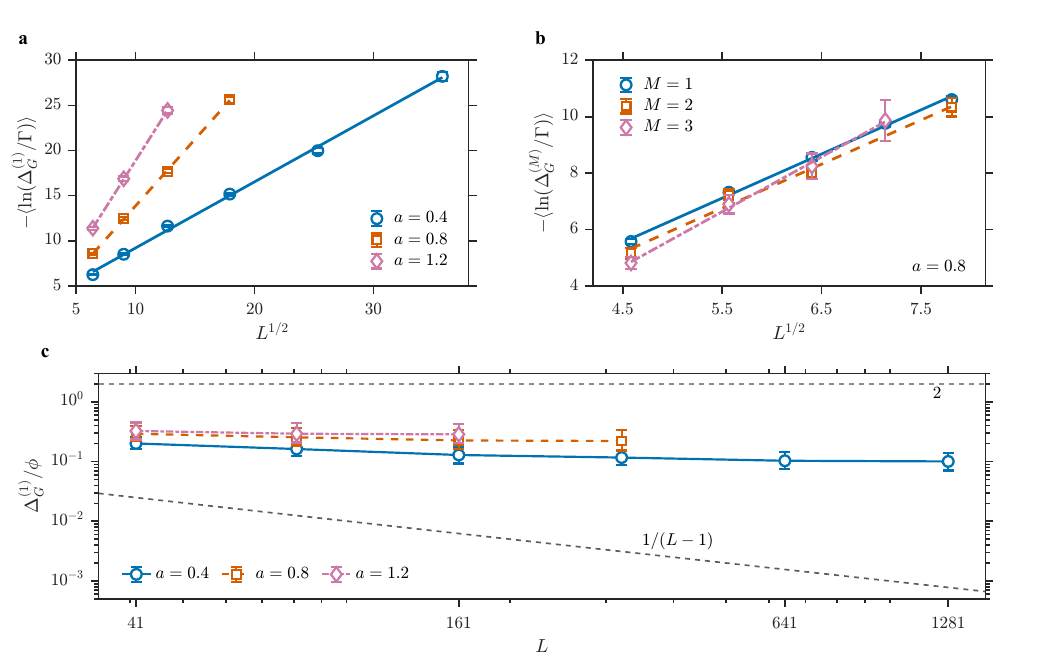}
\caption{\textbf{Random-walk barriers set the activated one-particle relaxation scale.}
\textbf{a}, Mean \(-\log(\Delta_G^{(1)}/\Gamma)\) at \(a=0.4,0.8,1.2\); lines are free-intercept linear fits in \(\sqrt L\).
\textbf{b}, Fixed-\(M\) spectra at \(a=0.8\), for \(M=1,2,3\). Bars in a and b are standard errors of the logarithmic gap.
\textbf{c}, Geometric mean of the paired ratio \(\Delta_G^{(1)}/\phi\), with interquartile bars. Dashed curves are the exact bounds \(1/(L-1)\) and 2. Every sample, not only the average, obeys these bounds. All chains have reflecting boundaries. The one-particle large-size windows contain 6,000 samples, and the common few-particle window contains 2,500, 320 and 120 samples for \(M=1,2,3\), respectively. Individual sizes and counts are specified in Methods. All rates are in units of \(\Gamma\).}
\label{fig:gaps}
\end{figure}

More precisely, let \(b(x)\) be a standard Brownian bridge on \([0,1]\). The linearly interpolated conditioned profile \(\Phi/(2a\sqrt{L-1})\) converges to \(b\)~\cite{Liggett1968Bridge}. The barrier functional is continuous in the uniform norm, while the bond term and logarithmic slack in Eq.~\eqref{eq:barrierbound} vanish on this scale. Hence
\begin{equation}
\frac{-\log(\Delta_G^{(1)}/\Gamma)}{2a\sqrt{L-1}}
\ \Longrightarrow\
\sup_{0\le x\le1}\left[\min\left(\sup_{u\le x}b(u),\sup_{u\ge x}b(u)\right)-b(x)\right],
\label{eq:bridge_limit}
\end{equation}
where the arrow denotes convergence in distribution (Supplementary Note~4). This fixes the fluctuating activated scale, rather than a deterministic coefficient fitted to a finite size window.

The one-particle ensembles follow this trend in Fig.~\ref{fig:gaps}a, and the paired samplewise ratio in Fig.~\ref{fig:gaps}c directly tests Eq.~\eqref{eq:pathbound}. Sturm-bracketed, arbitrary-precision eigenvalues retain even the slowest samples. Supplementary Fig.~2a,b gives the clean comparison and longer-chain conductance; Supplementary Fig.~2c reports finite-window gap slopes without imposing equality with conductance slopes. The \(M=2,3\) spectra in Fig.~\ref{fig:gaps}b support a similar trend at accessible sizes, but the path proof is one-particle only. At fixed density, a configuration-space bottleneck need not reduce to one spatial cut, and its thermodynamic scaling remains undetermined.
\FloatBarrier

\subsection*{Experimental implementation and observable signatures}

A candidate platform is a gas of fermionic \(^{40}\mathrm{K}\) atoms in an optical lattice, with two addressable hyperfine states encoding \(\uparrow,\downarrow\). Strong transverse confinement would isolate one-dimensional chains; the longitudinal lattice depth sets the reciprocal tunneling \(t\), while the scattering length and confinement set the repulsion \(U\). A finite segment with one atom per site and suppressed tunneling at its ends realizes half filling and reflecting boundaries. Single-atom-resolved potassium imaging~\cite{Cheuk2015}, occupation-dependent correlated spin-flip tunneling~\cite{Xu2018} and spatially resolved optical addressing~\cite{Zheng2022} supply separate precedents for preparation, control and readout. Doublon-decay measurements resolve charge lifetimes~\cite{Strohmaier2010}, and theoretical studies address site-selective charge dynamics in driven Hubbard superlattices~\cite{Cheng2024Site}. None of these results demonstrates the conditional return network required here.

Each bond requires four number-conserving return channels: either oriented doublon--hole defect must return to either of the two singly occupied spin configurations. A proposed realization couples each source to short-lived auxiliary two-atom states with distinguishable environmental decay records. On resonance, the auxiliary scheme in Methods gives a partial rate \(|\Omega_{j,s\nu}|^2/\kappa_A\), where \(\Omega_{j,s\nu}/2\) is the conditional coupling and \(\kappa_A\) the auxiliary decay rate. Adjusting the two intensities at fixed sum changes \(p_j^{R,L}\) while keeping the source linewidth \(\kappa_0\) fixed; both source orientations must have the same final-state probabilities. The loss readout in Ref.~\cite{Xu2018} cannot replace return to a trapped pair. A suitable atomic level scheme must conserve atom number and encoded magnetization, distinguish all four channels and suppress unwanted motional excitation. This network remains a proposal. Spatially uncorrelated differential light-shift noise could suppress Mott coherences; a uniform field cannot distinguish configurations of the same magnetization.

A concrete starting point for calibration is the locally tested parameter set \(U/t=10\), \(\kappa_0/t=200\), \(\gamma_\phi/t=1\) and \(a_j=\pm0.8\), where local noise is represented by jumps \(\sqrt{\gamma_\phi}S_j^z\). Equal numbers of positive and negative bonds impose the balanced condition. On a positive bond, \(p_j^R\simeq0.832\) and \(p_j^L\simeq0.168\); a negative bond interchanges them. These choices give \(\Gamma/t\simeq0.0148\), with charge decay much faster than spin dephasing and exchange. Full three-site calculations support the reduction on finite exchange times (Methods and Supplementary Note~5), not an experimental realization of these rates. The auxiliary decay must further satisfy \(\kappa_A\gg\kappa_0\), and the lattice band spacing must accommodate the return-induced broadening without appreciable excitation out of the intended motional manifold. These constraints and measured heating and loss determine whether an absolute optical parameter set is usable.

Calibration would begin in isolated double wells. Preparing each oriented defect and measuring its survival and final spin outcomes determines its linewidth and branching probabilities. Preparing \(|L_j\rangle\) and \(|R_j\rangle\) instead gives the two transfer curves. Their slopes after charge and spin-coherence transients, but before substantial population redistribution, determine \(W_j^{R,L}\); the microscopic slope at zero time is not the effective rate. The resulting \(a_j=\tfrac12\log(W_j^R/W_j^L)\) and \(\Gamma_j=\sqrt{W_j^RW_j^L}\) reconstruct \(\Phi_j\) independently of the density measurement. This also tests whether the calibrated end-to-end bias vanishes and whether unequal bond activities must be retained.

For a chain measurement, prepare a singly occupied segment, write one down spin or a specified fixed-\(M\) configuration, and then apply the calibrated return channels and local noise. At each evolution time \(\tau\), freeze tunneling by raising the lattice depth, switch off the engineered dynamics and map the spin populations onto site-resolved occupation images. Independent preparations at each time, with calibrated spin-selective readout and separate charge measurements, yield \(n_j(\tau)\), atom-number survival and residual doublon density. The late-time profile tests Eq.~\eqref{eq:pss} using the independently reconstructed landscape. For \(M=1\), a spatial density imbalance with nonzero overlap on the slow mode probes \(\Delta_G^{(1)}\); several initial spin positions and readout regions distinguish the slow decay from a faster mode with larger measurement weight. Fits must be made within each fixed disorder sample before averaging their rates. Short balanced chains, for example \(L=9\), should precede a size-scaling study: at the test point the population model already gives a median relaxation time \(t/\Delta_G^{(1)}\simeq1.34\times10^3\), with a broad sample distribution. Observation must extend to the sample's relaxation time while remaining well below heating and loss times. Reciprocal rates, uniform bias, reordered balanced signs and sign reversal provide separate controls for diffusion, boundary accumulation, sample dependence and landscape reversal.

\section*{Discussion}

Erratic Liouvillian models already connect prescribed random rates to internal localization and Sinai-type transport~\cite{Longhi2026QST}. Here the rates follow from a number-conserving microscopic return process with an explicit domain of validity. Reciprocal hopping creates virtual charge defects; the reservoir chooses their final spin states, and suppression of the residual spin coherence converts that choice into asymmetric population exchange. The resulting hard-core filling differs from localization produced by a uniform field and asymmetric tunneling~\cite{Wang2022Tightly}, topological doublons in a driven non-Hermitian model~\cite{Wang2026Gapless}, or interaction-induced bosonic caging~\cite{Yang2025NonHermitian}. In the present resolved dynamics, exclusion and the local rate ratios alone determine the stationary measure, without an attractive binding mechanism.

The short-chain comparison supports the reduction over finite exchange times, not uniformly in system size or on activated timescales. The fixed-\(M\) stationary measure is exact for the transition-resolved population generator, while the localization data cover its stationary population and first four nonstationary physical right modes at the tested sizes. The conductance bound and Brownian-bridge limit apply to one particle. The \(M=2,3\) spectra support a dilute finite-size trend, but neither fixed-density thermodynamic relaxation nor the unrestricted compact-jump spectrum follows from that evidence.

Reservoir engineering uses dissipation to select quantum states and dynamics~\cite{Diehl2008,Verstraete2009}. The present construction places that control in the virtual intermediates of a constrained system. Return probabilities fix stationary weights; the corresponding bond activities also determine how rapidly particles cross a bottleneck. When the virtual channels are controllable and distinguishable, and residual coherence is suppressed on the required timescale, local return measurements can therefore predict both many-body accumulation and slow transport without altering the reciprocal Hamiltonian.

\section*{Methods}

\subsection*{Microscopic operators and fixed-total-linewidth return channels}

We describe the number-conserving reservoir and spin monitoring by a Markovian master equation~\cite{Lindblad1976,Gorini1976,Daley2014},
\begin{equation}
\dot\rho=\mathcal L\rho=-i[H_{\mathrm{Hub}},\rho]
+\sum_{j,s,\nu}\mathcal D[R_{j,s}^{\nu}]\rho
+\gamma_{\mathrm m}\sum_{\mathcal C\in P}\mathcal D[|\mathcal C\rangle\langle\mathcal C|]\rho,
\qquad \mathcal D[X]\rho=X\rho X^\dagger-\tfrac12\{X^\dagger X,\rho\}.
\label{eq:master}
\end{equation}
The Fock orbitals are ordered \((1\uparrow,1\downarrow,2\uparrow,2\downarrow,\ldots)\), with ascending creation operators and the first orbital encoded by the lowest bit. In the isolated-bond basis \((L,R,D_R,D_L)\),
\begin{equation}
H=\begin{pmatrix}0&0&t&t\\0&0&-t&-t\\t&-t&U&0\\t&-t&0&U\end{pmatrix}.
\label{eq:bondmatrix}
\end{equation}
Fermionic reset forms and their signs are given in Supplementary Note~1. Configuration-projector monitoring is the theoretical ideal used to derive the exactly resolved population block. It leaves both \(P\)-diagonal populations and the \(Q\rho Q\) block unchanged, damps distinct \(P\)-coherences by \(\gamma_{\mathrm m}\), and damps \(P\rho Q\) by \(\gamma_{\mathrm m}/2\). Consequently, the virtual coherence width is \((\kappa_0+\gamma_{\mathrm m})/2\), not just \(\kappa_0/2\). The latter is used in Eq.~\eqref{eq:coherent_effective} only within Eq.~\eqref{eq:hierarchy}.

Local \(S^z\) noise is a distinct microscopic dephasing model and changes Mott--charge coherence widths at finite rate. For \(L_j^\phi=\sqrt{\gamma_\phi}S_j^z\), a coherence between eigenconfigurations \(x,y\) decays at \(\gamma_\phi\sum_j(s_{jx}-s_{jy})^2/2\). An exchanged Mott pair has decay \(\gamma_\phi\), whereas its coherence with an adjacent doublon--hole state has extra decay \(\gamma_\phi/4\). Local noise and configuration-projector monitoring therefore have different charge broadening, even though they resolve the same bond-exchange coherence.

\subsection*{Charge elimination and Mott-coherence elimination}

Let \(V=QH_tP\). With charge propagation neglected to leading order, \(H_{\mathrm{NH}}^{-1}=(U-i\kappa_0/2)^{-1}\) on a newly created adjacent defect. Effective-operator elimination~\cite{Reiter2012,Kessler2012} gives \(B=-RH_{\mathrm{NH}}^{-1}V\) and \(H_{\mathrm{ex}}=-\tfrac12V^\dagger[H_{\mathrm{NH}}^{-1}+(H_{\mathrm{NH}}^{-1})^\dagger]V\), producing Eq.~\eqref{eq:coherent_effective}. For diagonal \(\rho_P\), \(\mathcal P_{\mathrm d}\mathcal L_{\mathrm{eff}}\mathcal P_{\mathrm d}\) has transition rate \(2\kappa_0p_j^\nu t^2/D_0\); the elastic diagonal contribution cancels between recycling and anticommutator. Off-diagonal terms of size \(J_{\mathrm{rec}}\) and \(W\) remain until the second elimination.

Eliminating these coherences produces corrections of order \((J_{\mathrm{rec}}+W_{\max})^2/\gamma_{\mathrm m}\) locally. The purely Hamiltonian two-state correction is reciprocal, but mixed coherent--dissipative corrections need not be. Monitoring also changes the leading charge rates by a relative amount of order \(\gamma_{\mathrm m}/\kappa_0\). These estimates motivate Eq.~\eqref{eq:hierarchy}, with fixed coordination and bounded bias. Higher-order charge propagation and separated-defect states are not assumed to share a uniform dissipative gap. The expansion concerns initially Mott states on the exchange timescale, with direct finite-size checks in Fig.~\ref{fig:projection}.

At finite monitoring rate, the Mott--charge coherence width changes the leading denominator. Supplementary Note~1 quantifies this correction and shows why strong dephasing alone does not justify the unbroadened charge denominator.

Fixed linewidth is sufficient but not necessary for the directional ratio. Suppose environmentally distinguishable charge sources \(s\) have total linewidths \(\kappa_s\), detunings \(U_s\), equal-magnitude couplings \(t_s\) from the two Mott states, and shared final-state probabilities \(p_j^\nu\). In the same two-stage limit,
\begin{equation}
W_j^\nu=p_j^\nu\sum_s\frac{\kappa_s|t_s|^2}{U_s^2+\kappa_s^2/4},\qquad
\frac{W_j^R}{W_j^L}=\frac{p_j^R}{p_j^L}.
\label{eq:factorization}
\end{equation}
Source-dependent lifetimes change the common activity but cancel from this ratio. Source-dependent final-state probabilities or unequal excitation magnitudes need not have this property (Supplementary Note~1).

\subsection*{Full Hubbard--Lindblad propagation and error measures}

The full generators are built in fixed \(N=L,N_\downarrow=M\), using exact fermionic signs. With column vectorization, \(\mathcal D[R]=R^*\otimes R-\tfrac12[I\otimes R^\dagger R+(R^\dagger R)^T\otimes I]\). We propagate the full density matrix, the charge-eliminated density matrix and \(G\) independently. Figure~\ref{fig:projection} uses seven hoppings \(t/E_0=0.2\times2^{-k}\), \(k=0,\ldots,6\), and 81 times over \(0\le\Gamma\tau\le4\). Its error is \(\max_{\tau,\mathcal C}|\langle\mathcal C|\rho_{\mathrm{full}}|\mathcal C\rangle-p_{\mathcal C}|\), with the corresponding definition for the coherent effective model. Trace, positivity, charge weight and eigenvector-conditioning checks are included in Source Data. Independent matrix-exponential checks are also supplied for the proposed operating point. Errors quoted are over the stated time grid.

The zero-frequency Schur complement \(K(0)=\mathcal L_{dd}-\mathcal L_{df}\mathcal L_{ff}^{-1}\mathcal L_{fd}\) is used as a local rate diagnostic, where \(d\) retains Mott populations and \(f\) all other matrix elements. It is the leading term of the frequency-dependent memory kernel, not an exact all-time Markov generator. Initial transients and the slow projected trace must be assessed by propagation, as above. No exchange rate is refitted.

\subsection*{Transition-resolved generator and physical right modes}

The off-diagonal entry \(G_{\eta',\eta}\) is the allowed exchange rate from \(\eta\) to \(\eta'\); each diagonal entry is minus its column's total escape rate. For one particle, \(A=-D^{-1/2}G D^{1/2}\) has diagonal escape rates and off-diagonals \(-\sqrt{W_j^RW_j^L}\). Physical right modes are reconstructed as \(r_q=D^{1/2}u_q\), not plotted as symmetric eigenvectors. We use 80-decimal-digit Sturm bracketing and inverse iteration for the first four nonstationary modes. Their eigen-equation residuals and zero-sum conditions are checked before conversion to plotting data. With \(w_{qj}=|r_q(j)|^2/\sum_j|r_q(j)|^2\), participation is \(1/\sum_jw_{qj}^2\) and width is \([\sum_jw_{qj}(j-\sum_\ell\ell w_{q\ell})^2]^{1/2}\). Figure~\ref{fig:overview}d averages these diagnostics over four modes within each sample, then over 80 independent balanced samples at each \(L=31,61,91\); bars are standard errors across samples. Uniform-bias and clean chains are deterministic controls. Supplementary Fig.~4a,b gives complementary peak diagnostics.

For the illustration in Fig.~\ref{fig:overview}c, we fix the middle ensemble size \(L=61\) before inspecting profiles. Each of its 80 samples is represented by its mean participation fraction and mean width divided by \(L\), using the first four nonstationary modes. We standardize each coordinate by the ensemble mean and sample standard deviation, then select the sample nearest the componentwise median in Euclidean distance, breaking ties by the lowest sample index. This selects sample 79, seed 20321910. Its spectrum is recalculated with the same precision and checked against the ensemble data; it remains counted once in that ensemble. Peak positions play no role in selection.

\subsection*{Exact canonical recursion}

For prefix polynomials \(E_m^{(\ell)}\) and suffix polynomials \(F_m^{(\ell)}\),
\begin{equation}
E_m^{(\ell)}=E_m^{(\ell-1)}+z_\ell E_{m-1}^{(\ell-1)},\qquad
n_j^{\mathrm{can}}=\frac{z_j}{Z_M}\sum_{m=0}^{M-1}E_m^{(j-1)}F_{M-1-m}^{(j+1)}.
\label{eq:dp}
\end{equation}
We use \(E_0=F_0=1\), impossible occupations zero, and logarithmic sum-exp arithmetic. Small-chain enumeration independently checks all fixed sectors. Waterlines use \(\mu=\log[n_j^{\mathrm{gc}}/(1-n_j^{\mathrm{gc}})]-\Phi_j\); the four displayed densities in Fig.~\ref{fig:filling} are rounded labels for integer \(M\).

\subsection*{Population gap, conductance and numerical evaluation}

We define \(\Delta_G^{(M)}=\min_{\lambda\ne0}[-\operatorname{Re}\lambda(G^{(M)})]\). A resolved-jump density-matrix realization has additional off-diagonal decay eigenvalues; a compact-jump or full Hubbard Liouvillian has still different blocks. Population and full density-matrix gaps are therefore labeled separately. Supplementary Note~2 gives a two-state counterexample to their automatic equality. A separate Mott-only compact-jump finite-size check is given in Supplementary Note~5 and Supplementary Fig.~3.

The displayed gap windows retain every predefined sample identity. They contain 6,000 one-particle samples in the large-size window and 2,500 in the window shared with the few-particle calculation. Single-particle gaps are bracketed using a tridiagonal Sturm count. Decimal precision is set from the landscape range and increased by 25 digits for a second calculation; relative brackets are narrower than \(10^{-14}\), and the two logarithmic gaps agree within \(10^{-8}\). No eigenvalue-floor censoring is applied. For \(M=2,3\), sparse symmetric shift-invert calculations at two shifts identify the known zero vector separately and require both the gap-relative residual and the inter-shift difference to be below \(10^{-4}\). These criteria retain all 440 few-particle samples.

The one-particle windows are \(L=41,81,161,321,641,1281\) at \(a=0.4\), \(41,81,161,321\) at 0.8, and \(41,81,161\) at 1.2. Each point has 500 samples, except \(L=641,1281\), which have 250. The common-window \(M=1\) sizes are \(21,31,41,51,61\), with 500 samples each. \(M=2\) uses \(21,31,41,61\), with 80 samples each; \(M=3\) uses \(21,31,41,51\), with 40,40,25,15 samples. Fits have free intercepts; slope errors propagate sample-mean variances through ordinary least squares. Omitting the smallest size gives a separate window-sensitivity diagnostic, not a new scaling law.

Conductance is evaluated with logarithmic prefix/suffix sums. Every spectral sample has a paired conductance and is checked against Eq.~\eqref{eq:pathbound}. A separate reproducible ensemble of 200 samples at each \(L=101,201,401,801,1601,3201\) and each disorder strength supplies Supplementary Fig.~2b. Supplementary Fig.~1b uses 200 common samples across densities; uncertainty in normalized concentration includes the shared denominator through paired bootstrap resampling. Seeds, complete bond sequences, raw observations, solver checks and fit windows accompany the source tables.

\subsection*{Finite-size sensitivity to calibrated rate errors}

Local elimination accuracy and stationary-profile accuracy are different conditions. Consider two positive nearest-neighbor exclusion generators with the same allowed moves and \(\widetilde W_j^{R,L}=W_j^{R,L}(1+e_j^{R,L})\), where \(|e_j^{R,L}|\le\epsilon<1\). The error in each logarithmic rate ratio is bounded by \(d_\epsilon=\log[(1+\epsilon)/(1-\epsilon)]\). Writing \(m=\min(M,L-M)\), the corresponding fixed-sector stationary measures satisfy
\begin{equation}
\|\widetilde P_{\mathrm{ss}}^{(M)}-P_{\mathrm{ss}}^{(M)}\|_{\mathrm{TV}}
\le\tanh\!\left[\frac{m(L-1)d_\epsilon}{4}\right],
\label{eq:stationary_sensitivity}
\end{equation}
where total variation is half the sum of absolute probability differences. This follows by bounding the oscillation of the configuration-weight error and normalizing the resulting exponential reweighting (Supplementary Note~6). It is a worst-case finite-size bound, not a claim that typical errors saturate it. Errors common to both rates on a bond leave their ratio, and hence the stationary measure, unchanged exactly.

For any two finite Markov generators \(G\) and \(\widetilde G=G+E\), contraction of stochastic semigroups gives \(\|e^{\widetilde G\tau}-e^{G\tau}\|_1\le\tau\|E\|_1\), with the induced column norm. At \(\tau\sim1/\Delta_G\), \(\|E\|_1/\Delta_G\ll1\) is thus a sufficient, not necessary, accuracy criterion. This can be much stricter than an error measured relative to a local exchange rate. These comparisons apply within a closed population description; extra charge modes, additional transitions or coherent memory in the full Hubbard model need separate estimates. Independent finite-state enumeration tests both bounds and the unchanged-ratio case.

\subsection*{Experimental calibration and validity conditions}

For each bond, defect orientation \(s\) and final state \(\nu\), an auxiliary two-atom state has Hamiltonian \(\delta_{j,s\nu}|A\rangle\langle A|+(\Omega_{j,s\nu}|A\rangle\langle D_s|/2+\mathrm{H.c.})\) and decay \(\sqrt{\kappa_A}|\nu\rangle\langle A|\). Here \(\delta_{j,s\nu}\) is the detuning and \(\Omega_{j,s\nu}/2\) is the coupling matrix element. For \(|\Omega_{j,s\nu}|\ll\sqrt{\delta_{j,s\nu}^2+(\kappa_A/2)^2}\), auxiliary elimination gives
\begin{equation}
\kappa_{j,s\nu}=\frac{\kappa_A|\Omega_{j,s\nu}|^2}{4[\delta_{j,s\nu}^2+(\kappa_A/2)^2]},
\label{eq:auxrate}
\end{equation}
and the source light shift \(-|\Omega_{j,s\nu}|^2\delta_{j,s\nu}/[4(\delta_{j,s\nu}^2+\kappa_A^2/4)]\). On resonance, \(|\Omega_{j,s\nu}|^2=\kappa_A\kappa_0p_j^\nu\), with \(\kappa_A\gg\kappa_0\). Orthogonal environmental outcomes are required for the four channels to add incoherently. Both atoms must remain trapped and preserve the encoded magnetization. This specifies an operator scheme, not a verified potassium laser sequence.

Spatially uncorrelated fluctuations of the differential hyperfine energy can supply local spin dephasing. A uniform field couples only to total magnetization and cannot distinguish exchanged Mott configurations. Quantum-Zeno control in \(^{87}\mathrm{Rb}\) supplies a precedent for measurement-controlled dynamics~\cite{Schafer2014}, not for the proposed potassium return network or for this specific local noise spectrum. Monitoring-induced charge broadening must be included separately, as above.

At the local-noise test point stated in Results, full three-site propagation gives a maximum population error below 0.019 and charge probability below \(6\times10^{-4}\) over \(0\le\Gamma\tau\le4\). The exchanged-spin coherence decays at \(\gamma_\phi\), while the extra charge-coherence width is \(\gamma_\phi/4\); it is not identical to configuration-projector monitoring. The larger and smaller target bond rates are approximately \(0.0330t\) and \(0.00665t\). The charge, spin-coherence and exchange times are consequently well separated locally, but this does not establish accuracy on the activated chain relaxation time. With \(\hbar=1\), all quoted times are in inverse hopping units. If spectroscopy gives the hopping energy as \(t/h=\nu_t\) in hertz, one inverse hopping unit is \((2\pi\nu_t)^{-1}\) seconds. Absolute optical rates require a specified level scheme, lattice spectrum and recoil-heating budget; the dimensionless test is not measured hardware performance.

Preparation of each oriented defect in an isolated double well calibrates its survival and final-state probabilities. Starting instead from a Mott pair determines exchange rates in the window \(\kappa_0^{-1},\gamma_{\mathrm{coh}}^{-1}\ll\tau\ll W_{\max}^{-1}\), where \(\gamma_{\mathrm{coh}}\) is the measured exchanged-spin coherence decay rate. The literal derivative at \(\tau=0\) is not the effective rate. Then \(a_j=\tfrac12\log(W_j^R/W_j^L)\) and \(\Gamma_j=\sqrt{W_j^RW_j^L}\) reconstruct the landscape without a spatial-profile fit. Unwanted atom loss, indistinguishable environmental outcomes and auxiliary light shifts are independent diagnostics. Long-time decay should be fitted within each disorder sample, with preparation/readout overlap checked; averaging heterogeneous exponentials need not reveal a single gap. Both heating and loss times must substantially exceed \(1/\Delta_G\).

The chain readout uses repeated preparations, not a sequence of destructive images of the same evolving state. Spin-selection and imaging errors are calibrated using known spin patterns; empty and doubly occupied sites require an independent charge-sensitive sequence rather than an assumption of unit filling. For one excitation, a useful signal is \(I(\tau)=\sum_jw_j[n_j(\tau)-n_j^{\mathrm{ss}}]\), where the fixed real weights \(w_j\) define a spatial imbalance. The measured rates determine a candidate slow right mode, allowing weights and initial spin positions with nonzero mode overlap to be chosen before fitting. A stable late-time decay rate across these choices is stronger evidence for \(\Delta_G^{(1)}\) than a fit to a single local trace. Fits should also be stable when the start of the fitting window is varied. Atom loss is measured separately; normalizing only surviving shots must not hide a change of dynamical sector. For fixed \(M>1\), the canonical-profile comparison remains distinct from the one-particle gap measurement.

For the test point above, 100 balanced samples at each size give median population-model relaxation times \(t\tau_r\simeq1.34\times10^3\) at \(L=9\) and \(2.38\times10^5\) at \(L=41\); the latter 95th percentile is about \(7.25\times10^7\) (Supplementary Table~1). This broad distribution favors short-chain profile tests before an activated-scaling experiment. Short-time agreement does not establish the microscopic gap at these later times. Supplementary Notes~5 and 6 specify the observation budget and finite-size rate-error bounds.

\section*{Data availability}

Source Data supplied with the manuscript contain all plotted numerical values, individual disorder samples and bond sequences, fit inputs, microscopic trajectories, and the precision and convergence records underlying the four main and four Supplementary figures.

\section*{Code availability}

The code used for the calculations and figures will be deposited in a public repository upon acceptance of the manuscript.

\section*{Acknowledgements}

X.Z.Z. acknowledges support from the National Natural Science Foundation of China under Grants No. 12675024 and No. 12275193. W.-M.L. acknowledges support from the National Key R\&D Program of China under Grants Nos. 2024YFF0726700, 2021YFA1400900, and 2021YFA0718300; the National Natural Science Foundation of China under Grants Nos. 12334012, 12234012, and 52327808; the Space Application System of China Manned Space Program; and the Elite Revitalizing Inner Mongolia Program (2025TGL05).

\section*{Author contributions}

Y.T.W. performed the calculations and prepared the figures. X.Z.Z. conceived and supervised the project. X.Z.Z. and W.M.L. developed the physical interpretation. All authors discussed the results and contributed to writing the manuscript.

\section*{Competing interests}

The authors declare no competing interests.

\bibliographystyle{naturemag}
\setlength{\bibsep}{2pt}
\bibliography{main_nc_refs}

\clearpage
\appendix
\setcounter{section}{0}
\setcounter{equation}{0}
\setcounter{figure}{0}
\setcounter{table}{0}
\renewcommand{\theequation}{S\arabic{equation}}
\renewcommand{\thefigure}{S\arabic{figure}}
\renewcommand{\thetable}{S\arabic{table}}
\setcounter{secnumdepth}{2}

\renewcommand{\topfraction}{0.9}
\renewcommand{\textfraction}{0.08}
\renewcommand{\floatpagefraction}{0.9}

\graphicspath{{figures/}}

\begin{center}
	{\large\bf Supplemental Material for ``Reservoir-conditioned virtual returns generate random Liouvillian skin localization in a reciprocal Mott insulator''}\par\vspace{0.6em}
	Y. T. Wang, X. Z. Zhang, and W. M. Liu\par\vspace{0.2em}
\end{center}

\vspace{0.8em}

\maketitle
\renewcommand{\theequation}{S\arabic{equation}}
\renewcommand{\figurename}{Supplementary Figure}
\renewcommand{\tablename}{Supplementary Table}

Note~1 derives the microscopic return network and its two-stage reduction. Note~2 defines the population generator and its physical right modes. Notes~3 and 4 derive the finite-density stationary measure and one-particle barrier bounds. Note~5 details the separate compact-jump diagnostic and the experimental proposal. Note~6 bounds the sensitivity of finite-size stationary measures and finite-time population dynamics to rate errors. Four supplementary figures accompany these notes.

\section*{Supplementary Note 1: Microscopic return network and two controlled eliminations}

\subsection*{Fock convention and number-conserving resets}

We order orbitals as \((1\uparrow,1\downarrow,2\uparrow,2\downarrow,\ldots)\), with ascending creation operators in a Fock state. Orbital \(1\uparrow\) is the lowest bit. The model has \(N=L\) atoms and fixed down-spin number \(M\). Inside the singly occupied manifold \(P\), the down-spin indicator is \(\eta_j=1/2-S_j^z\), and the bit \(2^{j-1}\) in the reduced spin basis denotes \(\eta_j=1\). This spin-bit basis is separate from the fermionic orbital-bit basis.

On a bond, write
\begin{equation}
	|L\rangle=|\downarrow,\uparrow\rangle,
	\quad |R\rangle=|\uparrow,\downarrow\rangle,
	\quad |D_R\rangle=|0,\uparrow\downarrow\rangle,
	\quad |D_L\rangle=|\uparrow\downarrow,0\rangle.
\end{equation}
Their fermionic bit states are \((6,9,12,3)\). For reciprocal hopping \(H_t=-t\sum_\sigma(c_{2\sigma}^\dagger c_{1\sigma}+\mathrm{H.c.})\), direct anticommutation gives
\begin{align}
	H_t|L\rangle&=t(|D_R\rangle+|D_L\rangle),&
	H_t|R\rangle&=-t(|D_R\rangle+|D_L\rangle),\\
	H_t|D_R\rangle&=t(|L\rangle-|R\rangle),&
	H_t|D_L\rangle&=t(|L\rangle-|R\rangle).
\end{align}
With charge energy \(U\), the bond matrix is
\begin{equation}
	H=\begin{pmatrix}0&0&t&t\\0&0&-t&-t\\t&-t&U&0\\t&-t&0&U\end{pmatrix}.
	\label{eq:S_bond}
\end{equation}
The relative minus sign cannot be discarded when calculating a coherent effective jump. There is no \(e^{\pm a_j/2}\) in the hopping matrix element.

Let \(e_j=(1-n_{j\uparrow})(1-n_{j\downarrow})\) and \(d_j=n_{j\uparrow}n_{j\downarrow}\) project onto empty and doubly occupied sites. Number-conserving local realizations of the four resets are
\begin{align}
	R_{j,R}^R&=\sqrt{\kappa_0p_j^R}\,c_{j\uparrow}^\dagger c_{j+1,\uparrow}e_jd_{j+1},&
	R_{j,R}^L&=\sqrt{\kappa_0p_j^L}\,c_{j\downarrow}^\dagger c_{j+1,\downarrow}e_jd_{j+1},\\
	R_{j,L}^R&=\sqrt{\kappa_0p_j^R}\,c_{j+1,\downarrow}^\dagger c_{j\downarrow}d_je_{j+1},&
	R_{j,L}^L&=\sqrt{\kappa_0p_j^L}\,c_{j+1,\uparrow}^\dagger c_{j\uparrow}d_je_{j+1},
	\label{eq:S_resets}
\end{align}
where \(p_j^{R,L}=e^{\pm a_j}/(2\cosh a_j)\). Projectors act first. These operators respectively map each source defect into the stated final Mott configuration, with possible overall fermionic phases that can be absorbed separately in each jump. Spectator sites are unchanged. All channels commute with total atom number and total \(S^z\). The source label \(s\) and final-state label \(\nu\) have different meanings. Orthogonal reservoir records make the corresponding environmental outcomes distinguishable and justify a sum of dissipators over both labels.

For a given source,
\begin{equation}
	\sum_\nu (R_{j,s}^\nu)^\dagger R_{j,s}^\nu
	=\kappa_0|D_{j,s}\rangle\langle D_{j,s}|.
	\label{eq:S_width}
\end{equation}
This identity fixes the defect population decay rate. The total master equation, including the recycling terms, is trace preserving. Pure dephasing \(\mathcal D[|D\rangle\langle D|]\) is also trace preserving, but does not reset a charge population. It cannot be used as a population-return channel merely because it broadens a coherence.

\subsection*{What the monitoring damps}

Configuration-projector monitoring is the theoretical ideal used to derive the exactly resolved population block. It is generated by
\begin{equation}
	\mathcal L_m=\gamma_m\sum_{\mathcal C\in P}\mathcal D[|\mathcal C\rangle\langle\mathcal C|].
\end{equation}
For matrix element \(|x\rangle\langle y|\), a diagonal Hermitian jump with eigenvalues \(\ell_x,\ell_y\) contributes \(-|\ell_x-\ell_y|^2/2\). Therefore \(\mathcal L_m\) vanishes on diagonal Mott populations and on \(Q\rho Q\), acts as \(-\gamma_m\) on distinct Mott coherences, and acts as \(-\gamma_m/2\) on \(P\rho Q\) and \(Q\rho P\). A coherence between a Mott state and an adjacent charge state has the decay rate
\begin{equation}
	g_{PQ}=\frac{\kappa_0+\gamma_m}{2}.
	\label{eq:S_gpq}
\end{equation}
The charge population lifetime is \(1/\kappa_0\), not \(1/g_{PQ}\). These quantities must not be interchanged.

If the noise is instead \(L_j^\phi=\sqrt{\gamma_\phi}S_j^z\), its coherence decay is
\begin{equation}
	\gamma_{xy}=\frac{\gamma_\phi}{2}\sum_j(s_{jx}-s_{jy})^2.
	\label{eq:S_localdeph}
\end{equation}
For \(|L\rangle,|R\rangle\), the squared differences sum to 2, so the decay is \(\gamma_\phi\). For either Mott state and an adjacent \(D_s\), they sum to \(1/2\), giving \(\gamma_\phi/4\). Thus \(g_{PQ}=\kappa_0/2+\gamma_\phi/4\) for this local realization. It is a distinct microscopic noise model, not an exact replacement of configuration monitoring at the same parameter value.

\subsection*{Charge elimination preserves elastic--exchange interference}

Set \(V=QH_tP\) and initially neglect monitoring corrections to the charge propagator. On the newly created one-defect states,
\begin{equation}
	V=t(|D_R\rangle+|D_L\rangle)(\langle L|-\langle R|),\qquad
	H_{NH}^{-1}=\frac{Q}{U-i\kappa_0/2}.
\end{equation}
The imaginary denominator represents amplitude damping at half the population decay rate. Equivalently,
\begin{equation}
	\int_0^\infty d\tau\,e^{-(\kappa_0/2+iU)\tau}
	=\frac{-i}{U-i\kappa_0/2}.
\end{equation}
The effective-operator construction~\cite{Reiter2012} gives
\begin{align}
	B_s^\nu&=-R_s^\nu H_{NH}^{-1}V
	=-\frac{t\sqrt{\kappa_0p_\nu}}{U-i\kappa_0/2}
	|\nu\rangle(\langle L|-\langle R|),\label{eq:S_B}\\
	H_{\mathrm{ex}}&=-\tfrac12V^\dagger[H_{NH}^{-1}+(H_{NH}^{-1})^\dagger]V
	=-\frac{2Ut^2}{D_0}(|L\rangle-|R\rangle)(\langle L|-\langle R|),
	\label{eq:S_Heff}
\end{align}
where \(D_0=U^2+(\kappa_0/2)^2\). The corresponding Mott density-matrix generator is
\begin{equation}
	\mathcal L_P\rho_P=-i[H_{\mathrm{ex}},\rho_P]+\sum_{s,\nu}\mathcal D[B_s^\nu]\rho_P+\mathcal L_m\rho_P.
	\label{eq:S_LP}
\end{equation}
For \(\nu=R\), \(B_s^R\) contains exchange \(|R\rangle\langle L|\) and elastic return \(-|R\rangle\langle R|\). Expanding \(\mathcal D[B_s^R]\) produces their cross terms. Only the source and final-state environmental outcomes are distinguishable; the two initial-state amplitudes within this same jump remain coherent.

At \(\gamma_m=0\), the triplet \(|T_0\rangle=(|L\rangle+|R\rangle)/\sqrt2\) satisfies \(V|T_0\rangle=B_s^\nu|T_0\rangle=H_{\mathrm{ex}}|T_0\rangle=0\). It is also dark in the full four-state model. Splitting \(B_s^\nu\) into independent elastic and exchange jumps removes their destructive interference and incorrectly destroys this dark state. Direct application of the full and effective operators confirms the cancellation.

For diagonal \(\rho_P=\sum_i p_i|i\rangle\langle i|\), the off-diagonal entries of its projected population generator are \(\sum_{s,\nu}|(B_s^\nu)_{fi}|^2\), \(f\ne i\). Each of the two sources contributes \(\kappa_0p_Rt^2/D_0\) to \(L\to R\), giving
\begin{equation}
	W_R=\frac{2\kappa_0p_Rt^2}{D_0},\qquad
	W_L=\frac{2\kappa_0p_Lt^2}{D_0},\qquad
	W_{R,L}=\Gamma_j e^{\pm a_j},\quad
	\Gamma_j=\frac{\kappa_0t^2}{D_0\cosh a_j}.
	\label{eq:S_rates}
\end{equation}
The diagonal elastic term cancels from populations between recycling and anticommutator, but its coherent cross terms do not vanish as operators.

\subsection*{Second elimination and a nonempty joint hierarchy}

Let \(\mathcal P_d\) retain Mott diagonal elements and \(\mathcal Q_d\) its coherences. Write \(\mathcal V\) for the coherent-exchange and effective dissipative terms in Eq.~\eqref{eq:S_LP}. Since \(\mathcal L_m=-\gamma_m\) on distinct Mott coherences, their feedback has leading structure
\begin{equation}
	\delta G=-\mathcal P_d\mathcal V\mathcal Q_d
	(\mathcal Q_d\mathcal L_m\mathcal Q_d)^{-1}
	\mathcal Q_d\mathcal V\mathcal P_d.
\end{equation}
At fixed coordination its local scale is at most of order \((J_{\mathrm{rec}}+W_{\max})^2/\gamma_m\), where \(J_{\mathrm{rec}}=2Ut^2/D_0\) is the off-diagonal pair-exchange matrix element; the conventional Heisenberg coupling is \(2J_{\mathrm{rec}}\). The purely Hamiltonian two-state contribution can be obtained explicitly. For coherence \(c=\rho_{LR}\), splitting \(\delta E\) and Hamiltonian exchange \(J\),
\begin{equation}
	\dot c=-(\gamma_{\mathrm{coh}}+i\delta E)c-iJ(p_R-p_L),\qquad
	c\simeq-\frac{iJ(p_R-p_L)}{\gamma_{\mathrm{coh}}+i\delta E}.
\end{equation}
Substitution into \(\dot p_L=-2J\operatorname{Im}c\) gives the reciprocal correction \(2J^2\gamma_{\mathrm{coh}}/(\gamma_{\mathrm{coh}}^2+\delta E^2)\). This formula alone does not describe the mixed terms of Eq.~\eqref{eq:S_LP}.

The sufficient conditions used in the main text are
\begin{equation}
	\epsilon_{ch}=t/\sqrt{D_0}\ll1,\quad \gamma_m/\kappa_0\ll1,
	\quad W_{\max}/\gamma_m\ll1,
	\quad (J_{\mathrm{rec}}+W_{\max})^2/(\gamma_m W_{\min})\ll1.
	\label{eq:S_hierarchy}
\end{equation}
The second condition controls monitoring-induced virtual broadening; the last controls relative population-rate errors and is stronger than simply \(\gamma_m\gg J_{\mathrm{rec}}\) in some parameter regimes. For fixed \(U,\kappa_0\), finite bias and fixed local coordination, let \(t=\epsilon E_0\) and \(\gamma_m=\epsilon E_0\). Both \(J_{\mathrm{rec}}\) and \(W\) are \(O(\epsilon^2 E_0)\), so all dimensionless conditions vanish with \(\epsilon\). More generally \(\gamma_m/E_0=\epsilon^\alpha\) with \(0<\alpha<2\) works. Unbounded bias, growing coordination or a size-dependent observation time require a separate estimate.

On a chain, the first hop from \(P\) creates an adjacent defect. Propagation within \(Q\) and creation of further defects require additional hoppings and enter beyond the local leading order. Once separated, however, a defect can decay slowly because the reset acts only on adjacent pairs. The local denominator therefore does not imply a uniform gap of the entire \(Q\)-Liouvillian or convergence of its slowest mode. The reduction describes the leading spin dynamics from an initially Mott state; full short-chain propagation retains separated configurations and tests this description on finite exchange times.

\subsection*{Finite monitoring and the zero-frequency diagnostic}

One can expose the broadening correction without an effective-operator approximation. At fixed \(\gamma_m>0\) and \(t\to0\), the coherence of one charge source with a Mott initial state satisfies
\begin{equation}
	\dot\rho_{D,i}=-(g_{PQ}+iU)\rho_{D,i}-iV_{Di}(p_i-p_D)+O(t\rho_{LR}).
\end{equation}
Solving the fast coherence gives a transfer into that source at \(2t^2g_{PQ}p_i/(U^2+g_{PQ}^2)\), to leading order in its small population. Balancing this transfer with decay \(\kappa_0p_D\), multiplying by the final-state probability, and summing the two sources gives
\begin{equation}
	W_R=\frac{4p_Rt^2g_{PQ}}{U^2+g_{PQ}^2}+o(t^2),\qquad
	g_{PQ}=\frac{\kappa_0+\gamma_m}{2}.
	\label{eq:S_broadened_rate}
\end{equation}
This local formula has fixed nonzero monitoring as its asymptotic assumption. It is not asserted for every chain or arbitrary noise model. In the joint hierarchy it reduces to Eq.~\eqref{eq:S_rates}.

For a complete finite Liouvillian, separate Mott populations \(p\) from other entries \(q\): \(\dot p=Ap+Bq\), \(\dot q=Cp+Dq\). Laplace elimination gives
\begin{equation}
	K(z)=A+B(z-D)^{-1}C=K(0)-zBD^{-2}C+\cdots,
	\qquad K(0)=A-BD^{-1}C.
	\label{eq:S_memory}
\end{equation}
There is also an initial-layer term \(B(z-D)^{-1}q(0)\). The frequency derivative changes the slow-coordinate normalization, so \(K(0)\) is a diagnostic rather than an exact all-time master equation for normalized Mott probabilities. The fast block must be invertible where this diagnostic is applied. We therefore assess the initial layer and memory effects by direct propagation, without refitting the exchange rate.

For \(t=0.01,U=10,\kappa_0=2,a=0.8,\gamma_m=1\), the unbroadened rate is \(3.29512\times10^{-6}\), whereas the exact zero-frequency four-state diagnostic gives \(4.88300\times10^{-6}\). Equation~\eqref{eq:S_broadened_rate} accounts for the leading difference. At \(\gamma_m=0\) the triplet dark state must instead be retained; one cannot obtain a unique classical exchange merely by deleting its coherence. The full monitoring scan is included in Source Data.

Main-text Fig.~2a,b compares full and reduced dynamics for \(L=3,M=1\); Fig.~2c,d gives population errors and charge admixture for \(L=2,M=1\) and \(L=3,M=1\), retaining all 4 and 9 Fock states, respectively. We use \(U/E_0=10\), \(\kappa_0/E_0=2\), \(\gamma_m=t\), and \(t/E_0=0.2\times2^{-k}\) for \(k=0,\ldots,6\). The three-site bonds are \((0.8,-0.8)\), while the two-site bond is 0.8. Initially all atoms are singly occupied and the down spin is at site 1. Source Data contain unnormalized projected populations, charge probability, trace, positivity and zero-frequency diagnostics. The convergence test uses 81 uniformly spaced times over \(0\le\Gamma\tau\le4\), not all later activated times in an arbitrarily long chain.

\subsection*{Equal linewidth is sufficient, not necessary}

Suppose source \(s\) has detuning \(U_s\), total decay \(\kappa_s\), and equal excitation magnitudes \(|t_{s,L}|=|t_{s,R}|=|t_s|\). If all sources share final-state probabilities \(p_R,p_L\), leading resolved rates factorize as
\begin{equation}
	W_R=p_R\sum_s\frac{\kappa_s|t_s|^2}{U_s^2+(\kappa_s/2)^2},\qquad
	W_L=p_L\sum_s\frac{\kappa_s|t_s|^2}{U_s^2+(\kappa_s/2)^2}.
	\label{eq:S_factorization}
\end{equation}
Hence unequal source lifetimes can change the common activity without changing \(W_R/W_L=p_R/p_L\). The equal-width construction makes this activity uniform for binary bias, but is not required by the directional mechanism. Factorization can fail if final-state selection is source dependent or the two initial states have unequal excitation magnitudes. Directional linewidth corrections must then be calculated rather than inferred from a linewidth mismatch alone. Spatial variation of a positive \(\Gamma_j\) leaves the canonical stationary measure unchanged but affects kinetics.

\section*{Supplementary Note 2: Population generator and physical right modes}

\subsection*{Configuration space and gap conventions}

The fixed-\(M\) state space is \(\Omega_{L,M}=\{\boldsymbol\eta:\eta_j=0,1,\sum_j\eta_j=M\}\). The column \(\boldsymbol\eta\) of \(G\) lists rates out of that configuration: \(G_{\boldsymbol\eta',\boldsymbol\eta}=W_j^R\) for \(10\to01\), \(W_j^L\) for \(01\to10\), and its diagonal is minus their sum. Thus \(\boldsymbol1^TG=0\). In spin notation, the right move is \(S_j^+S_{j+1}^-\). This convention agrees with the fermionic states in Note~1 and the arrows in main-text Fig.~1.

A resolved quantum embedding can use \(J_{\eta',\eta}=\sqrt{G_{\eta',\eta}}|\eta'\rangle\langle\eta|\) for each allowed move. Its diagonal subspace is governed by \(G\), but its off-diagonal matrix units decay at \(-(k_\eta+k_{\eta'})/2\), where \(k_\eta\) is the escape rate. Additional monitoring and resolved elastic returns add coherence decay. For two states with bidirectional rates \(\Gamma\), \(G\) has nonzero decay \(2\Gamma\), whereas the jump-only quantum embedding has coherence decays \(\Gamma\). Consequently
\begin{equation}
	\Delta_G^{(M)}=\min_{\lambda(G)\ne0}[-\operatorname{Re}\lambda(G)]
\end{equation}
is labeled as a population gap throughout. Its equality to the gap of an unrestricted Liouvillian requires a separate block comparison. The compact and microscopic spectra cannot be assigned this equality by notation.

\subsection*{Detailed balance and sample dependence}

Set \(\Phi_1=0\), \(\Phi_{j+1}-\Phi_j=\log(W_j^R/W_j^L)\). For \(M=1\), \(\pi_j=e^{\Phi_j}/Z_1\) satisfies \(\pi_jW_j^R=\pi_{j+1}W_j^L\). A reflecting chain is reversible for any positive rates, even if \(\Phi_L\ne\Phi_1\). For balanced binary disorder, there are equally many \(+a\) and \(-a\) bonds, so \(L\) is odd and the cumulative walk is conditioned to return to zero. This is a bridge, not an unconstrained spatial random walk.

The end-to-end logarithmic rate bias \(\mathcal B_{\mathrm{end}}=\Phi_L-\Phi_1\) vanishes for the balanced bridge. It is not a thermodynamic cycle affinity, since the open chain has no nontrivial spatial cycle. Local rates can still differ, \(W_j^R-W_j^L\ne0\), and transient currents need not vanish. A finite bridge can have endpoint maxima, ties or secondary basins. Reversing \(a_j\) changes \(\Phi_j\) to \(-\Phi_j\); densities and mode shapes respond through their nonlinear dependence on that profile. Balance constrains the disorder, not each finite-time trajectory.

For stationary weights \(D_{\eta\eta}=P_{ss}^{(M)}(\eta)\), detailed balance makes
\begin{equation}
	A=-D^{-1/2}GD^{1/2}=A^T,\qquad
	G(D^{1/2}u_q)=-\lambda_qD^{1/2}u_q.
\end{equation}
The physical right mode is \(r_q=D^{1/2}u_q\); the symmetric eigenvector alone is not the physical population mode. In one dimension the random-potential and rate-reweighting descriptions are mathematically equivalent. Random skin localization refers here to the concentration of the physical modes generated by asymmetric rates. Reversibility is retained, different modes may peak at different sites, and spatial pinning alone does not establish a point-gap invariant.

\subsection*{Mode diagnostics and independent controls}

Each mode is normalized through \(w_{qj}=|r_q(j)|^2/\sum_\ell|r_q(\ell)|^2\). We record
\begin{equation}
	\mathrm{PR}_q=\frac{1}{\sum_jw_{qj}^2},\quad
	\bar j_q=\sum_jjw_{qj},\quad
	\sigma_q^2=\sum_j(j-\bar j_q)^2w_{qj},\quad
	\mathcal P_q=\frac1L\#\{j:\Phi_j\le\Phi_{j_q}\},
\end{equation}
where \(j_q\) is the first maximum of \(|r_q|\). Participation measures occupied support, whereas width also detects separated peaks. The percentile \(\mathcal P_q\) is insufficient alone: it equals one everywhere in the flat clean landscape.

The ensemble contains 80 independent balanced samples at each \(L=31,61,91\), \(a=0.8\), generated by shuffling equal numbers of positive and negative bond biases. Seeds are \(20260831+1000L+s\), \(s=0,\ldots,79\); Source Data provide the complete bond sequences, so reproducing a sample does not require regenerating its random numbers. The first four nonstationary modes are computed at 80 decimal digits using Sturm-bracketed eigenvalues and shifted tridiagonal inverse iteration. We require the gap-relative eigen-equation residual and the physical zero-sum residual to be below \(10^{-14}\). Modes are averaged within each sample before ensemble means and standard errors; four modes from one landscape are not counted as four independent disorder realizations.

For main-text Fig.~1c we fix \(L=61\), the middle ensemble size, before inspecting mode profiles. For each of its 80 samples, form the pair \((\overline{\mathrm{PR}}/L,\overline\sigma/L)\), averaging the first four nonstationary modes. Standardize each coordinate by its ensemble mean and sample standard deviation, and select the sample with the smallest Euclidean distance to the componentwise median, breaking ties by sample index. This rule selects sample 79, seed 20321910. No peak-location or landscape-percentile criterion enters the selection. Its four modes are recalculated at the same precision and agree with their ensemble diagnostics within \(10^{-12}\). The sample is already part of the ensemble and is counted only once. Source Data supply all candidate distances, the selected sign string and the plotted profiles; Supplementary Software implements the selection.

Main-text Fig.~1d displays participation and width. Supplementary Fig.~\ref{fig:S_modes}a reports the mean peak percentile, including the clean counterexample in which every site has percentile one. Supplementary Fig.~\ref{fig:S_modes}b gives the fraction of stationary profiles whose global maximum includes an endpoint. These diagnostics describe finite-size internal localization of the sampled first four nonstationary modes; neither supports a claim about an extensive fraction of the spectrum.

\section*{Supplementary Note 3: Exact stationary measure and finite-density filling}

\subsection*{Detailed balance in every fixed sector}

For an allowed move \(\eta\to\eta'\) from \(j\) to \(j+1\), the exponent changes by \(\Phi_{j+1}-\Phi_j\). Therefore
\begin{equation}
	\frac{P_{ss}^{(M)}(\eta')}{P_{ss}^{(M)}(\eta)}
	=e^{\Phi_{j+1}-\Phi_j}=\frac{W_j^R}{W_j^L},\qquad
	P_{ss}^{(M)}(\eta)W_j^R=P_{ss}^{(M)}(\eta')W_j^L.
\end{equation}
Summing pairwise cancellations gives \(GP_{ss}=0\), with
\begin{equation}
	P_{ss}^{(M)}(\eta)=Z_M^{-1}\exp\!\left(\sum_j\eta_j\Phi_j\right)\delta_{\sum_j\eta_j,M},\qquad
	Z_M=\sum_{x_1<\cdots<x_M}e^{\Phi_{x_1}+\cdots+\Phi_{x_M}}.
	\label{eq:S_canonical}
\end{equation}
Nearest-neighbor moves connect all fixed-\(M\) configurations when rates are positive, proving uniqueness for \(0<M<L\); the endpoint sectors contain one state. Any positive bond activity \(\Gamma_j\) cancels from the ratio and hence from this stationary measure. This does not make relaxation independent of \(\Gamma_j\).

The measure is a product weight conditioned on particle number. Its hard-core constraint, \(\eta_j\le1\), enforces single-site exclusion rather than attraction. At finite size, occupation varies smoothly with the weights; deterministic filling of the most favorable sites requires an additional large-contrast limit. Favorable regions need not be connected.

\subsection*{Log-domain dynamic programming}

Let \(z_j=e^{\Phi_j}\), \(E_m^{(\ell)}=e_m(z_1,\ldots,z_\ell)\), and \(F_m^{(\ell)}=e_m(z_\ell,\ldots,z_L)\). Partition the prefix subsets by whether site \(\ell\) is occupied:
\begin{equation}
	E_m^{(\ell)}=E_m^{(\ell-1)}+z_\ell E_{m-1}^{(\ell-1)},\quad
	E_0^{(\ell)}=1,\quad E_m^{(0)}=0\ (m>0).
\end{equation}
An analogous backward recursion gives \(F\). Fixing site \(j\) occupied leaves \(M-1\) particles to distribute on its left and right, so
\begin{equation}
	n_j^{can}=\frac{z_j}{Z_M}\sum_{m=0}^{M-1}E_m^{(j-1)}F_{M-1-m}^{(j+1)},\qquad Z_M=E_M^{(L)}.
	\label{eq:S_density}
\end{equation}
Prefix and suffix construction and all site sums cost \(O(LM)\) time and storage. Numerical implementation stores logarithms. A two-term addition uses \(\operatorname{logaddexp}(x,y)=\max(x,y)+\log[1+e^{-|x-y|}]\), with impossible occupations represented by \(-\infty\). This prevents exponential overflow and avoids subtracting nearly equal partition functions. We verify \(0\le n_j\le1\), \(\sum_jn_j=M\), and agreement with complete enumeration for \(L=5,7,9\), all nontrivial \(M\).

Introducing a fugacity \(e^\mu\) gives \(\prod_j(1+e^\mu z_j)\). Differentiation yields
\begin{equation}
	n_j^{gc}=\frac{1}{1+e^{-(\Phi_j+\mu)}},\qquad \sum_jn_j^{gc}=M.
\end{equation}
The sum is strictly increasing in \(\mu\), so a bracketed scalar root determines it. Its waterline is \(-\mu\), and the inverse relation is \(\mu=\log[n_j^{gc}/(1-n_j^{gc})]-\Phi_j\). We compare the canonical and grand-canonical densities using \(\varepsilon_L=L^{-1}\sum_j|n_j^{can}-n_j^{gc}|\). The size scan is a finite-sample diagnostic, not a uniform ensemble-equivalence theorem for all random landscapes.

\subsection*{Filling statistics and uncertainty}

A puddle is a contiguous run of sites with \(n_j^{can}>1/2\). We measure the largest run divided by \(L\), and the density concentration
\begin{equation}
	C(\rho)=L\sum_j(n_j^{can}/M)^2.
\end{equation}
The plotted normalized concentration is the ratio of disorder means \(\langle C(\rho)\rangle/\langle C(\rho_0)\rangle\), \(\rho_0\simeq0.01\), not the mean of per-sample ratios. All densities use the same 200 landscapes. Its uncertainty is obtained by resampling the sample index jointly in numerator and denominator (2,000 bootstrap resamples, seed 82913); the largest-run bars are standard errors. Threshold dependence is a geometric diagnostic, not evidence for a thermodynamic phase.

\begin{figure}[!htbp]
	\centering
	\includegraphics[width=\textwidth]{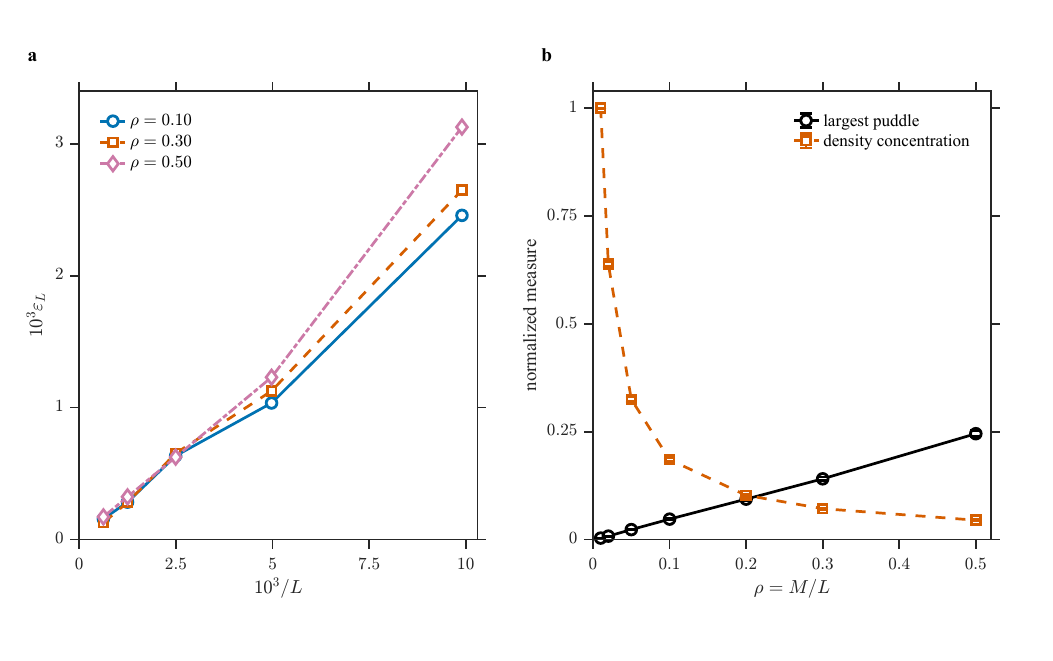}
	\caption{\textbf{Canonical filling checks and sample-averaged geometry.}
		\textbf{a}, Mean sitewise canonical--grand-canonical difference for balanced \(a=0.8\) profiles, using seed 13 at each size and integer particle numbers closest to the indicated fillings. The complete bond sequences are supplied in Source Data. Lines connect finite-size observations and are not asymptotic fits.
		\textbf{b}, Largest threshold-defined puddle fraction and normalized density concentration at \(L=401,a=0.8\), averaged over 200 common balanced samples. The latter includes shared-denominator uncertainty by paired bootstrap; integer fillings and seeds are in Source Data.}
	\label{fig:S_filling}
\end{figure}

\section*{Supplementary Note 4: Path conductance, activated relaxation and numerical spectra}

\subsection*{Why a spatial cut minimizes conductance on a path}

For one particle define \(\pi_j=e^{\Phi_j}/Z_1\), \(c_j=\pi_jW_j^R\), and \(m_j=\sum_{\ell\le j}\pi_\ell\). Let
\begin{equation}
	h=\min_j\frac{c_j}{\min(m_j,1-m_j)}.
\end{equation}
The usual conductance is \(\phi=\min_{0<\pi(A)\le1/2}Q(A,A^c)/\pi(A)\). Taking the smaller side of any prefix cut shows \(\phi\le h\). For the reverse inequality, decompose any allowed \(A\) into disjoint intervals. Its flow-to-mass ratio is a mass-weighted average of interval ratios. It therefore suffices to bound one interval of mass \(m\le1/2\). If the outside masses on its left and right are \(x,y\), then \(x+y=1-m\). Its two boundary flows satisfy
\begin{equation}
	c_{left}+c_{right}\ge h[\min(x,1-x)+\min(y,1-y)]\ge hm.
\end{equation}
The last inequality follows directly: if \(x,y\le1/2\), the bracket is \(1-m\ge m\); if \(x>1/2\), it is \(m+2y\ge m\), and similarly for \(y>1/2\). At an endpoint the absent flow and outside mass vanish. Every interval ratio is at least \(h\), proving \(\phi\ge h\). Hence
\begin{equation}
	\phi=h=\min_j\frac{c_j}{\min(m_j,1-m_j)}.
	\label{eq:S_phi}
\end{equation}
This equality uses the path geometry; it is not assumed for the many-particle configuration graph.

\subsection*{Gap bound with only polynomial slack}

For a reversible population generator, the gap is the minimum Rayleigh quotient
\begin{equation}
	\Delta_G^{(1)}=\inf_{f:\operatorname{Var}_\pi f>0}
	\frac{\sum_{j=1}^{L-1}c_j(f_{j+1}-f_j)^2}{\operatorname{Var}_\pi f}.
\end{equation}
Using the indicator of a prefix with mass \(m\), this quotient is \(c/[m(1-m)]\le2c/\min(m,1-m)\). The minimum cut thus gives \(\Delta_G^{(1)}\le2\phi\). For the lower bound, write
\begin{align}
	\operatorname{Var}_\pi f
	&=\sum_{x<y}\pi_x\pi_y(f_y-f_x)^2\\
	&\le (L-1)\sum_{j=1}^{L-1}m_j(1-m_j)(f_{j+1}-f_j)^2\\
	&\le\frac{L-1}{\phi}\sum_{j=1}^{L-1}c_j(f_{j+1}-f_j)^2.
\end{align}
The first inequality is Cauchy--Schwarz along the path from \(x\) to \(y\); the second uses \(c_j\ge\phi\min(m_j,1-m_j)\ge\phi m_j(1-m_j)\). Combining the two variational bounds yields
\begin{equation}
	\frac{\phi}{L-1}\le\Delta_G^{(1)}\le2\phi.
	\label{eq:S_pathbound}
\end{equation}
Generic conductance inequalities give a lower bound quadratic in conductance~\cite{Cheeger1970,JerrumSinclair1989,LevinPeresWilmer2009}. The path geometry instead gives a bound linear in \(\phi\), with only the factor \(L-1\). This distinction fixes the exponential relaxation scale without assuming equality of finite-window slopes.

\subsection*{Logarithmic partition sums and the barrier functional}

With uniform \(\Gamma\), define \(Z_j^- =\sum_{\ell\le j}e^{\Phi_\ell}\) and \(Z_j^+ =\sum_{\ell>j}e^{\Phi_\ell}\). The full normalization cancels in Eq.~\eqref{eq:S_phi}:
\begin{equation}
	\frac{\phi}{\Gamma}=\min_j\frac{e^{\Phi_j+a_j}}{\min(Z_j^-,Z_j^+)},\qquad
	Y_\phi=-\log(\phi/\Gamma)=\max_j[\min(\log Z_j^-,\log Z_j^+)-\Phi_j-a_j].
\end{equation}
For any partial sum, its logarithm lies between the largest exponent and that exponent plus \(\log L\). Hence, with
\begin{equation}
	\mathcal B_L=\max_j[\min(\max_{\ell\le j}\Phi_\ell,\max_{\ell>j}\Phi_\ell)-\Phi_j-a_j],
\end{equation}
one has \(\mathcal B_L\le Y_\phi\le\mathcal B_L+\log L\). Equation~\eqref{eq:S_pathbound} then gives
\begin{equation}
	\mathcal B_L-\log2\le -\log(\Delta_G^{(1)}/\Gamma)
	\le\mathcal B_L+\log L+\log(L-1).
	\label{eq:S_barrierbound}
\end{equation}
The barrier is two-sided because a cut with negligible probability on one side need not control equilibration between populated basins. The total range \(\max\Phi-\min\Phi\) is not automatically this bottleneck.

For uniformly shuffled balanced \(\pm a\) bonds, the rescaled interpolated profile \(\Phi(x)/(2a\sqrt{L-1})\) converges in distribution to a standard Brownian bridge \(b(x)\)~\cite{Liggett1968Bridge}. Maxima over intervals and a final supremum are continuous in the uniform norm. The bounded bond term \(a_j\) and the logarithmic slack in Eq.~\eqref{eq:S_barrierbound} disappear on this scale. Consequently the one-particle logarithmic gap has the same limiting barrier functional
\begin{equation}
	\frac{-\log(\Delta_G^{(1)}/\Gamma)}{2a\sqrt{L-1}}
	\ \Longrightarrow\
	\sup_{0\le x\le1}\left[\min\left(\sup_{u\le x}b(u),\sup_{u\ge x}b(u)\right)-b(x)\right].
	\label{eq:S_bridge_limit}
\end{equation}
The limiting barrier fluctuates between samples, so the result does not assign a universal deterministic exponential coefficient. It is consistent with the Sinai random-environment mechanism~\cite{Sinai1982,BouchaudGeorges1990}. The path proof does not extend directly to arbitrary fixed \(M>1\) or fixed density.

\subsection*{High-precision one-particle eigenvalues}

The positive symmetric operator \(A=-D^{-1/2}GD^{1/2}\) has diagonal entries equal to escape rates. In units \(\Gamma=1\), its off-diagonals are \(-1\) for the binary model. Its zero vector is \(\sqrt\pi\), and its second eigenvalue is the gap. If \(d_j\) is the diagonal, the Sturm pivots are
\begin{equation}
	q_1=d_1-x,\qquad q_j=d_j-x-1/q_{j-1}.
\end{equation}
The count of negative pivots gives the number of eigenvalues below trial \(x\). Bisection brackets the second eigenvalue without subtracting a near-zero stationary eigenvalue from a double-precision result. All exponential coefficients are formed directly in decimal arithmetic, not converted from rounded double-precision matrices.

Precision begins at \(\max[50,\lceil(\max\Phi-\min\Phi)/\log10\rceil+35]\) decimal digits and is repeated with 25 additional digits. The relative bracket width is at most \(10^{-14}\), and the logarithmic gaps must agree within \(10^{-8}\). The full ensemble passes. Source Data include both bracket endpoints, precision, iterations, seed, bond string and the independent higher-precision recomputation. Extremely slow samples are retained; no floor-based deletion enters an ensemble mean. Clean analytical gaps and independent small dense diagonalizations are separate tests.

For fixed \(M=2,3\), we build the symmetric exclusion matrix directly on the \(\binom LM\) configurations. A permitted exchange contributes \(-\Gamma\) off diagonal and the appropriate escape rate to the diagonal. The exact zero vector is proportional to \(\exp[\tfrac12\sum_j\eta_j\Phi_j]\). Sparse shift-invert solves at shifts \(-10^{-8}\) and \(-10^{-10}\) require agreement within \(10^{-4}\) relative to the positive gap and a gap-relative residual below \(10^{-4}\). The zero vector is identified by overlap, not by a numerical threshold alone. All 320 two-particle and 120 three-particle samples are retained.

\subsection*{Samples, fits and limitations}

One-particle main-window seeds are \(20260521+100000L+1000\operatorname{round}(10a)+s\), with sample index \(s\) preserved from the Source Data. The common-window one-particle base is 20260524, and the \(M=2,3\) bases are \(20260520+M\). Complete bond strings are authoritative. The sample counts and sizes are listed in main Methods. A separate conductance ensemble uses base 20260831, the same size/amplitude offsets, and \(s=0,\ldots,199\). No conductance value depends on a fitted gap.

\begin{figure}[!htbp]
	\centering
	\includegraphics[width=\textwidth]{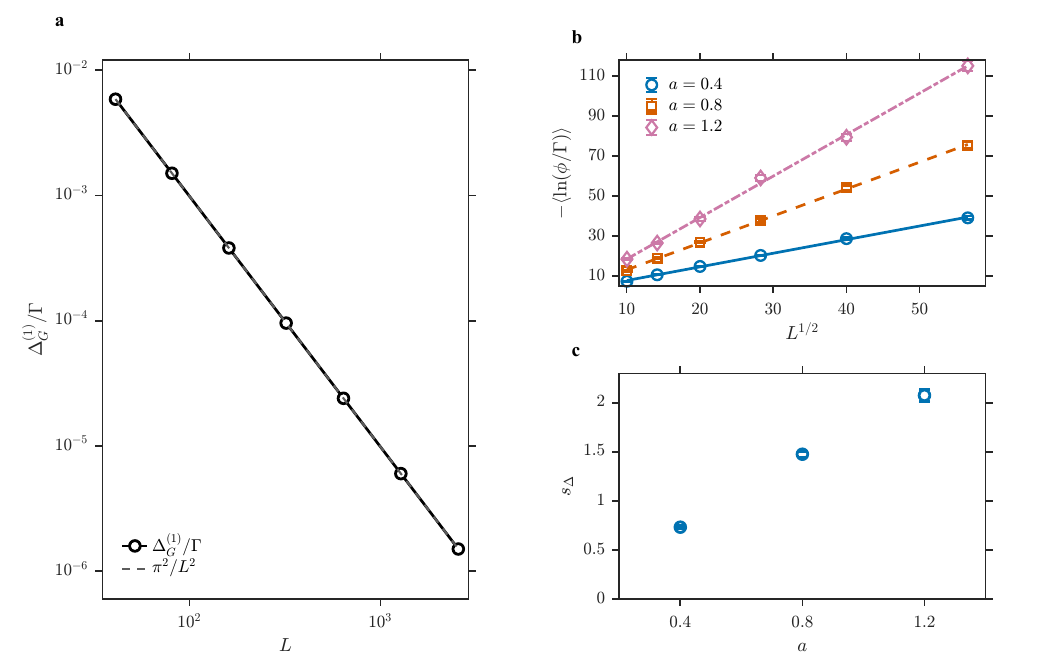}
	\caption{\textbf{Clean diffusion, longer-chain conductance and finite-window gap slopes.}
		\textbf{a}, Clean reflecting one-particle gap compared with \(\pi^2\Gamma/L^2\); the exact finite-size result is \(2\Gamma[1-\cos(\pi/L)]\).
		\textbf{b}, Mean \(-\log(\phi/\Gamma)\) from log-domain spatial cuts, for \(a=0.4,0.8,1.2\) and 200 independent samples per point at \(L=101,201,401,801,1601,3201\). Bars are standard errors; lines are free-intercept finite-window fits.
		\textbf{c}, Slopes of the one-particle logarithmic gaps in main-text Fig.~4a, with one-standard-error uncertainties propagated from independent sample means. No proportionality through the origin is imposed. All panels use reflecting chains. Different fitting windows do not imply equality of fitted gap and conductance slopes; the exact comparison is the samplewise bound in main-text Fig.~4c.}
	\label{fig:S_relax}
\end{figure}

For sample logarithms \(Y_s\), displayed means have standard error \(\operatorname{sd}(Y_s)/\sqrt N\). For a free-intercept fit with design matrix \(X=(1,\sqrt L)\), let \(B=(X^TX)^{-1}X^T\). Statistical coefficient covariance is \(B\operatorname{diag}(\mathrm{SE}_i^2)B^T\). This uncertainty does not include asymptotic-model error. We separately report the slope after removing the smallest size. Finite-window one-particle slopes are approximately \(0.732,1.476,2.076\) for \(a=0.4,0.8,1.2\), rather than imposing a fit through the origin. Supplementary Fig.~\ref{fig:S_relax}a,b gives the clean gap and longer-chain conductance, and Supplementary Fig.~\ref{fig:S_relax}c shows the gap slopes. Conductance windows differ and need not give the same finite-size slopes. The paired ratio in main-text Fig.~4c and Eq.~\eqref{eq:S_pathbound} are the direct common-scale test.

The \(M=2,3\) data are finite-size evidence of dilute slowing only. At finite density, the stationary landscape remains exactly known, but the controlling cut lies in configuration space and can change with exclusion. Neither puddle filling nor a local coherent-continuity check determines this thermodynamic relaxation. The population gap and the gap of an unrestricted Liouvillian also remain distinct, as explained in Note~2.

\section*{Supplementary Note 5: Compact-jump diagnostic and experimental details}

\subsection*{Compact density-matrix model}

\begin{figure}[!htbp]
	\centering
	\includegraphics[width=\textwidth]{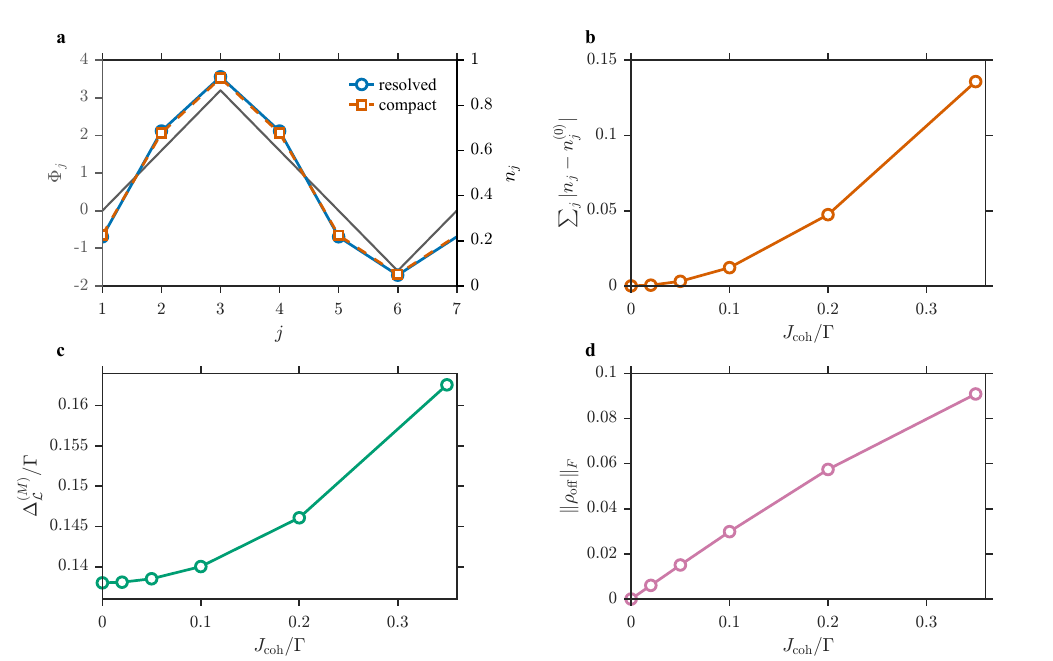}
	\caption{\textbf{A separate compact-jump model is continuous near its population limit.}
		\textbf{a}, Predefined landscape \(\Phi_j\) (left axis) and resolved/compact densities (right axis), the latter at \(J_{coh}/\Gamma=0.20\).
		\textbf{b}, Density distance from the canonical population measure.
		\textbf{c}, Gap of the compact-jump Liouvillian in units of \(\Gamma\).
		\textbf{d}, Off-diagonal stationary Frobenius norm. The calculation uses one \(L=7,M=3,a=0.8\) sample and \(\gamma_\phi=0.5\Gamma\). This finite-size Mott-only comparison tests continuity near the population limit. It does not establish the microscopic Hubbard elimination, an extensive-spectrum localization claim or fixed-density gap scaling.}
	\label{fig:S_compact}
\end{figure}

\begin{figure}[!htbp]
	\centering
	\includegraphics[width=\textwidth]{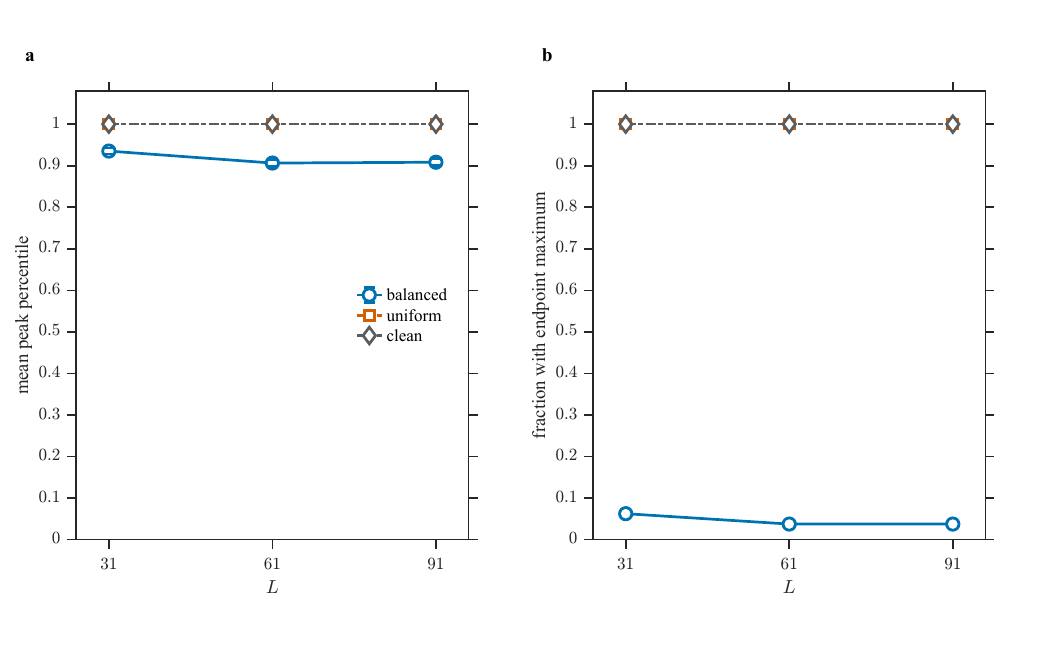}
	\caption{\textbf{Peak diagnostics complement the participation and width in main-text Fig.~1d.}
		\textbf{a}, Mean landscape percentile at the first maximum of each of the first four nonstationary physical right modes, averaged within each sample before the ensemble mean. A flat clean profile has percentile one everywhere, so this quantity alone does not establish localization.
		\textbf{b}, Fraction of samples whose stationary landscape maximum includes an endpoint, including ties. Bars are standard errors across 80 independent balanced samples at each \(L=31,61,91\), with \(a=0.8\) and reflecting boundaries. Clean and uniform-bias chains are deterministic controls. No extensive-spectrum conclusion is inferred from four modes.}
	\label{fig:S_modes}
\end{figure}

The independent Mott-only model has
\begin{equation}
	\widetilde L_j^R=\sqrt{W_j^R}S_j^+S_{j+1}^-,\quad
	\widetilde L_j^L=\sqrt{W_j^L}S_j^-S_{j+1}^+,\quad
	H_{coh}=J_{coh}\sum_j(S_j^+S_{j+1}^-+\mathrm{H.c.}),\quad
	L_j^\phi=\sqrt{\gamma_\phi}S_j^z.
\end{equation}
Each compact jump acts on all configurations with the same local bond pattern, but omits the elastic parts of \(B_s^\nu\). This separate Mott-only model tests continuity near population dynamics at finite size; it does not test microscopic Hubbard elimination, localization of an extensive spectral fraction or fixed-density gap scaling. We use the fixed \(L=7,M=3,a=0.8\) landscape and \(\gamma_\phi=0.5\Gamma\), scanning \(J_{coh}/\Gamma=0,0.02,0.05,0.1,0.2,0.35\). The Hilbert dimension is 35 and the Liouvillian dimension 1225. Supplementary Fig.~\ref{fig:S_compact}a,b compares stationary profiles and their deviation; Supplementary Fig.~\ref{fig:S_compact}c,d shows the compact Liouvillian gap and residual coherence.

With column vectorization, the coherent term is \(-i(I\otimes H-H^T\otimes I)\), and each dissipator is \(L^*\otimes L-\tfrac12[I\otimes L^\dagger L+(L^\dagger L)^T\otimes I]\). A trace-row replacement gives the stationary density matrix; sparse near-zero eigenvalues give its full compact Liouvillian gap. We check stationary residual, unit trace and positivity, and calculate \(n_j=\operatorname{Tr}(\rho\eta_j)\), \(\sum_j|n_j-n_j^{(0)}|\) and \(\|\rho_{off}\|_F\). These data are independently regenerated from the predefined landscape, not digitized from the figure.

\subsection*{An explicit auxiliary return network}

The proposed platform uses two hyperfine states of \(^{40}\mathrm{K}\) as spin labels in a one-dimensional optical lattice. The main text cites potassium microscopy, occupation-dependent correlated spin-flip tunneling and site-resolved addressing separately. The proposed network below is an additional theoretical requirement, not a demonstrated number-conserving potassium reset. In particular, the loss readout of the correlated-tunneling experiment cannot replace return to a trapped two-atom state.

For every bond, source defect and final state, introduce an auxiliary two-atom state \(|A_{j,s}^{\nu}\rangle\) with the same total atom number and magnetization as the initial pair. A conditional coupling and a return jump are
\begin{equation}
	H_A=\delta|A\rangle\langle A|+
	\tfrac12(\Omega|A\rangle\langle D_s|+\Omega^*|D_s\rangle\langle A|),\qquad
	L_A=\sqrt{\kappa_A}|\nu\rangle\langle A|.
\end{equation}
The auxiliary non-Hermitian energy is \(\delta-i\kappa_A/2\). For \(|\Omega|\ll\sqrt{\delta^2+\kappa_A^2/4}\), eliminating \(|A\rangle\) gives
\begin{align}
	R_{s\nu}^{eff}&=-\frac{\sqrt{\kappa_A}\Omega}{2(\delta-i\kappa_A/2)}|\nu\rangle\langle D_s|,\\
	\kappa_{s\nu}&=\frac{\kappa_A|\Omega|^2}{4(\delta^2+\kappa_A^2/4)},\qquad
	\delta U_s=-\frac{|\Omega|^2\delta}{4(\delta^2+\kappa_A^2/4)}.
\end{align}
At \(\delta=0\), the leading light shift vanishes and \(\kappa_{s\nu}=|\Omega_{s\nu}|^2/\kappa_A\). Taking \(|\Omega_{s\nu}|^2=\kappa_A\kappa_0p_\nu\) realizes the desired return channel, provided \(\kappa_A\gg\kappa_0\). Orthogonal emitted-photon states or auxiliary reservoir records are required; otherwise cross-channel coherent combinations change the effective operators. Both decay and coherent return paths must terminate in the trapped two-atom manifold. A lossy state that ejects an atom fails the number-conservation requirement.

Occupation-dependent Raman transfer, spatial addressing and repumping are possible ingredients for these operations. A concrete implementation still has to identify auxiliary states, suppress unwanted channels, resolve adjacent bonds, control recoil and off-resonant light shifts, and conserve the encoded magnetization. Local \(S^z\) noise supplies an operator-level monitoring alternative whose widths are given in Eq.~\eqref{eq:S_localdeph}. Its stochastic fields must resolve sites: a single spatially uniform field couples to total \(S^z\) and cannot distinguish configurations in the same magnetization sector. The noise spectrum must be broad on the slow exchange scale while avoiding appreciable population transfer to unwanted internal or motional states. The charge-coherence broadening remains part of the rate calibration.

\subsection*{Dimensionless operating point and lifetime budget}

An independently tested point for local spin noise is
\[
U/t=10,\qquad \kappa_0/t=200,\qquad \gamma_\phi/t=1,\qquad a=0.8.
\]
Here \(\epsilon_{ch}\simeq0.00995\), \(\Gamma/t\simeq0.014806\), \(J_{rec}/t\simeq0.0019802\), and the extra virtual width is \(\gamma_\phi/4\). For the balanced three-site pair of bonds, the predicted population gap is \(\Delta_G^{(1)}/t\simeq0.0329512\). Thus local exchange and relaxation times are about \(67.5/t\) and \(30.3/t\), respectively; the three-site gap can exceed the geometric-mean bond rate. Full microscopic propagation gives a maximum absolute projected-population deviation of 0.0187 and charge probability below \(5.6\times10^{-4}\) on the stated grid \(0\le\Gamma\tau\le4\). Matrix-exponential cross-checks agree to better than \(10^{-10}\) in trajectory norm. These are dimensionless model calculations, not measured optical performance.

The fast charge and exchanged-spin coherence times at this point are \(0.005/t\) and \(1/t\). To estimate the later observation window independently of the short-chain benchmark, we compute 100 balanced samples at each \(L=9,17,41\), \(a=0.8\). Supplementary Table~\ref{tab:S_budget} gives the resulting population-model relaxation times \(\tau_r=1/\Delta_G^{(1)}\) in hopping units.
\begin{table}[htbp]
	\centering
	\caption{\textbf{Sample-dependent observation times.} Median, 95th percentile and largest relaxation time among 100 balanced samples at each size, with \(a=0.8\) and \(\Gamma/t=0.01480594\). These statistics refer to the population generator, not to a long-chain microscopic simulation.}
	\label{tab:S_budget}
	\begin{tabular}{crrr}
		\toprule
		\(L\) & Median \(t\tau_r\) & 95th percentile \(t\tau_r\) & Maximum \(t\tau_r\)\\
		\midrule
		9  &\(1.34\times10^3\)&\(7.49\times10^3\)&\(1.02\times10^4\)\\
		17 &\(7.68\times10^3\)&\(6.83\times10^4\)&\(2.92\times10^5\)\\
		41 &\(2.38\times10^5\)&\(7.25\times10^7\)&\(5.34\times10^{10}\)\\
		\bottomrule
	\end{tabular}
\end{table}
These statistics describe the sampled population models; they are neither bounds nor predictions from a full microscopic long-time calculation. Source Data contain the bond arrays and population gaps, with seeds \(20260831+10000L+s\), \(s=0,\ldots,99\). The broad tail means that an observation time estimated from the mean exchange rate can miss the slowest samples.

The auxiliary-network limit further requires \(\kappa_A\gg\kappa_0\), selective addressing bandwidth below unwanted transitions, and negligible excitation of other lattice bands. Strong charge resetting suppresses the target exchange, so larger \(\kappa_0\) is not unconditionally beneficial. We do not assign absolute optical rates without a specified level scheme and band structure. Whatever scale \(t\) is realized, useful observation requires
\begin{equation}
	\tau_{loss},\tau_{heat}\gg\Delta_G^{-1},\qquad
	\Delta_G\tau_{life}\gg1,\quad \tau_{life}=\min(\tau_{loss},\tau_{heat}).
\end{equation}
Because one-particle relaxation is activated, a large-chain asymptotic regime may lie beyond this budget even when the local elimination works. Small chains and moderate disorder should therefore precede a scaling test. The dimensionless point above is an example of a locally consistent model, not an experimentally established universal window.

\subsection*{Preparation, calibration and measurement}

First prepare a half-filled segment with fixed magnetization, verifying single occupation by separate charge measurements and suppressing tunneling at both ends. Addressed spin rotations prepare a single down spin or a chosen multi-spin configuration. The same bond pattern must be reproduced across all preparations used for one time trace. On isolated double wells, prepare each oriented doublon--hole defect, measure its survival curve and record the two final Mott outcomes. Survival should be measured for total charge occupation as well as source orientation, so coherent redistribution among charge states is not mistaken for irreversible return. In the conditional-return calibration, coherent Hubbard transfer can be suppressed by raising the barrier, with return channels recalibrated for any induced level shifts. This determines source linewidths and source-conditioned return probabilities before inferring an effective bias. Agreement of the two source-conditioned branching ratios is a separate requirement from agreement of their lifetimes.

Prepare the singly occupied Mott pairs by spin rotations that preserve atom number. The literal transfer slope at \(\tau=0\) vanishes for a coherent microscopic excitation and is not the Markov exchange rate. Determine the coarse-grained transfer within
\begin{equation}
	\kappa_0^{-1},\gamma_{coh}^{-1}\ll\tau\ll W_{max}^{-1},
	\qquad a_j=\tfrac12\log(W_j^R/W_j^L),\quad \Gamma_j=\sqrt{W_j^RW_j^L}.
\end{equation}
Independent local calibrations reconstruct \(\Phi_j\) and test spatially uniform activity. Linewidth mismatch alone need not change the directional ratio if Eq.~\eqref{eq:S_factorization} applies; inconsistent final-state probabilities and spin-dependent excitation amplitudes must be checked separately.

At each chosen evolution time, freeze tunneling, turn off the return drive and noise, and read out the spin occupations. Each image comes from a fresh preparation. Spin-selective mapping or removal followed by occupation imaging requires calibration on known spin configurations, including state-transfer errors and detection loss. Separate charge-sensitive runs measure holes and doublons, which cannot in general be distinguished from singly occupied sites by an uncalibrated fluorescence count. Repeated snapshots determine the stationary profile for comparison with Eq.~\eqref{eq:S_density}, without fitting a second landscape.

For one-particle relaxation, form \(I(\tau)=\sum_jw_j[n_j(\tau)-n_j^{\mathrm{ss}}]\), with fixed real spatial weights \(w_j\) chosen to overlap the predicted slow right mode. A left--right density difference is one possibility, but is not guaranteed to have appreciable overlap in every sample. Use several initial spin positions and readout regions, and vary the lower end of the late-time fitting window. A rate that persists under these changes can be compared with the independently calibrated population gap. An observable orthogonal to the slow mode measures a faster rate. Averaging heterogeneous exponentials before fitting can obscure the gap; samplewise fits should precede disorder statistics. Record atom-number survival for the full ensemble: normalizing surviving shots alone can conceal loss-induced departures from the fixed sector. Heating, single-atom loss and transitions out of the Mott manifold must be negligible over the fitted interval, rather than interpreted as intrinsic slow relaxation.

Four controls separate the relevant mechanisms. Setting \(a_j=0\) tests reciprocal diffusion; a uniform sign tests boundary-directed accumulation; reordering a balanced multiset relocates extrema without changing its endpoint sum; \(a_j\to-a_j\) reverses the calibrated landscape. Charge population and spin coherence test the separation of timescales, auxiliary fluorescence resolves the return channels, and atom-number survival and heating constrain the observation window. The proposed measurements specify how to test an implementation; the present calculations do not demonstrate the required atomic network.

\section*{Supplementary Note 6: Finite-size sensitivity to rate imperfections}

The local elimination conditions do not alone bound the error of a long-chain stationary profile or its exponentially slow relaxation. A simple comparison within the population description quantifies this distinction. Suppose a second, positive nearest-neighbor exclusion generator has rates
\begin{equation}
	\widetilde W_j^{R,L}=W_j^{R,L}(1+e_j^{R,L}),\qquad |e_j^{R,L}|\le\epsilon<1.
\end{equation}
These assumptions exclude extra transitions and coherent memory. Its accumulated profile obeys
\begin{equation}
	|\delta\Phi_{j+1}-\delta\Phi_j|\le d_\epsilon,
	\quad d_\epsilon=\log\frac{1+\epsilon}{1-\epsilon},
	\quad \operatorname{osc}(\delta\Phi)\le(L-1)d_\epsilon,
\end{equation}
where \(\delta\Phi=\widetilde\Phi-\Phi\) and \(\operatorname{osc}f=\max f-\min f\). For fixed \(M\), let \(S(\eta)=\sum_j\eta_j\delta\Phi_j\). Two configurations differ on at most \(m=\min(M,L-M)\) occupied sites in each direction. Hence \(\operatorname{osc}S\le m\operatorname{osc}(\delta\Phi)\equiv R\), and their stationary measures are related by \(\widetilde P=P e^S/\langle e^S\rangle_P\).

To bound the total variation distance, set \(Y=e^S\in[y_-,y_+]\) and \(z=\langle Y\rangle_P\). Convexity of \(|Y-z|\) bounds it by its chord between the endpoints. Averaging this chord gives
\begin{equation}
	\begin{aligned}
		\|\widetilde P-P\|_{\rm TV}
		&=\frac{\langle|Y-z|\rangle_P}{2z}
		\le\frac{(y_+-z)(z-y_-)}{z(y_+-y_-)}\\
		&\le\frac{\sqrt{y_+}-\sqrt{y_-}}{\sqrt{y_+}+\sqrt{y_-}}
		\le\tanh\frac{m(L-1)d_\epsilon}{4}.
	\end{aligned}
	\label{eq:S_stationary_sensitivity}
\end{equation}
The middle maximum occurs at \(z=\sqrt{y_+y_-}\); constant \(S\) gives identical measures directly. Thus small bondwise relative errors need not give a size-independent stationary error. Exact preservation of each rate ratio is a special case with \(\delta\Phi=0\): then the stationary measure is unchanged even when bond activities change.

For two finite Markov generators \(G\) and \(\widetilde G=G+E\), their stochastic semigroups contract the induced column \(1\)-norm. Duhamel's identity therefore yields
\begin{equation}
	\|e^{\widetilde G\tau}-e^{G\tau}\|_1
	\le\int_0^\tau\|e^{\widetilde G(\tau-s)}\|_1\|E\|_1\|e^{Gs}\|_1ds
	\le\tau\|E\|_1.
	\label{eq:S_time_sensitivity}
\end{equation}
At \(\tau\sim\Delta_G^{-1}\), \(\|E\|_1/\Delta_G\ll1\) is a sufficient, not necessary, accuracy condition. It can be much stronger than a condition measured in local exchange units. Neither bound promotes the complete microscopic Hubbard model to a Markov population process: corrections involving extra charge modes or memory require their own estimates. These comparisons explain why the finite-exchange-time microscopic test and the exact activated gap of \(G\) are stated separately.

\end{document}